\pdfoutput=1
\documentclass[journal]{vgtc}                

\usepackage[utf8]{inputenc}

\ifpdf
  \pdfoutput=1\relax                   
  \usepackage{graphicx}                
  \DeclareGraphicsExtensions{.pdf,.png,.jpg,.jpeg} 
\else
  \usepackage{graphicx}                
  \DeclareGraphicsExtensions{.eps}     
\fi%

\graphicspath{{figures/}{pictures/}{images/}{./}} 

\usepackage{microtype}                 
\PassOptionsToPackage{warn}{textcomp}  
\usepackage{textcomp}                  
\usepackage{mathptmx}                  
\usepackage{times}                     
\usepackage{cite}                      
\usepackage{tabu}                      
\usepackage{booktabs}                  

\onlineid{0}

\vgtccategory{Research}
\vgtcpapertype{algorithm/technique}

\usepackage{cellspace}

\usepackage{ifpdf}

\ifpdf
  \pdfoutput=1\relax                   
  \usepackage{graphicx}                
  \DeclareGraphicsExtensions{.pdf,.png,.jpg,.jpeg} 
\else
  \usepackage{graphicx}                
  \DeclareGraphicsExtensions{.eps}     
\fi%

\usepackage{tabularx}

\graphicspath{{figures/}{pictures/}{images/}{./}}

\usepackage{microtype}         
\PassOptionsToPackage{warn}{textcomp} 
\usepackage{textcomp}                 
\usepackage{mathptmx}                
\usepackage{times}                    
       
\usepackage{cite}                   
\usepackage{tabu}                   
\usepackage{booktabs}               
\usepackage[normalem]{ulem} 
\usepackage{xspace}
\usepackage{amsfonts}
\usepackage{amsmath}
\usepackage{multirow}  

\usepackage[utf8]{inputenc}
\usepackage[T1]{fontenc}
\usepackage{balance}
\usepackage[hyphens]{url}
\usepackage{tabularx}

\usepackage{wrapfig,epsf}

\usepackage{xcolor}

\usepackage[hidelinks]{hyperref}

\newcommand{\squishlist}{
	\begin{list}{$\bullet$}
		{ \setlength{\itemsep}{0pt}      \setlength{\parsep}{3pt}
			\setlength{\topsep}{3pt}       \setlength{\partopsep}{0pt}
			\setlength{\leftmargin}{1.5em} \setlength{\labelwidth}{1em}
			\setlength{\labelsep}{0.5em} } }
	
\newcommand{\squishlisttwo}{
  \begin{list}{$\bullet$}
    { \setlength{\itemsep}{0pt}    \setlength{\parsep}{0pt}
      \setlength{\topsep}{0pt}     \setlength{\partopsep}{0pt}
      \setlength{\leftmargin}{2em} \setlength{\labelwidth}{1.5em}
      \setlength{\labelsep}{0.5em} } }
  
\newcommand{\squishend}{	\end{list}  }

\newcommand{\ie}{\emph{i.e.}\xspace}
\newcommand{\eg}{\emph{e.g.}\xspace}

\newcommand{\myparagraph}[1]{\smallskip\noindent\textbf{#1}\xspace}

\newcommand{\figref}[1]{\hyperref[#1]{\textbf{Figure~\ref*{#1}}}}
\newcommand{\figsref}[1]{\hyperref[#1]{Figures~\ref*{#1}}}
\newcommand{\tabref}[1]{\hyperref[#1]{Tab.~\ref*{#1}}}
\newcommand{\secref}[1]{\hyperref[#1]{Sec.~\ref*{#1}}}
\newcommand{\algref}[1]{\hyperref[#1]{Alg.~\ref{#1}}}

\title{MoReVis: A Visual Summary for Spatiotemporal Moving Regions}

\author{Giovani~Valdrighi, Nivan~Ferreira~\textit{Member,~IEEE}, Jorge~Poco~\textit{Member,~IEEE}}

\authorfooter{
\item Giovani~Valdrighi is with Funda\c{c}\~ao Getulio Vargas, Brazil. E-mail: {giovani.valdrighi@fgv.edu.br}.
\item Nivan Ferreira is with Universidade Federal de Pernambuco, Brazil. E-mail: {nivan@cin.ufpe.br}
\item Jorge Poco is with Funda\c{c}\~ao Getulio Vargas, Brazil and Universidad Católica San Pablo, Perú. E-mail: {jorge.poco@fgv.br}.
}

\abstract{
Spatial and temporal interactions are central and fundamental in many activities in our world. A common problem faced when visualizing this type of data is how to provide an overview that helps users navigate efficiently.
Traditional approaches use coordinated views or 3D metaphors like the Space-time cube to tackle this problem. However, they suffer from overplotting and often lack spatial context, hindering data exploration. More recent techniques, such as \textit{MotionRugs}, propose compact temporal summaries based on 1D projection. While powerful, these techniques do not support the situation for which the spatial extent of the objects and their intersections is relevant, such as the analysis of surveillance videos or tracking weather storms. 
In this paper, we propose MoReVis, a visual overview of spatiotemporal data that considers the objects' spatial extent and strives to show spatial interactions among these objects by displaying spatial intersections. Like previous techniques, our method involves projecting the spatial coordinates to 1D to produce compact summaries. However, our solution's core consists of performing a layout optimization step that sets the size and positions of the visual marks on the summary to resemble the actual values on the original space. We also provide multiple interactive mechanisms to make interpreting the results more straightforward for the user. 
We perform an extensive experimental evaluation and usage scenarios. Moreover, we evaluated the usefulness of MoReVis in a study with 9 participants. The results point out the effectiveness and suitability of our method in representing different datasets compared to traditional techniques.
}
\keywords{Spatiotemporal visualization, spatial interactions, spatial abstraction.}

\newcommand{\figReference}{
\begin{figure}[t!]
    \centering
    \includegraphics[width=\columnwidth,trim={25pt 0 25pt 10pt}]{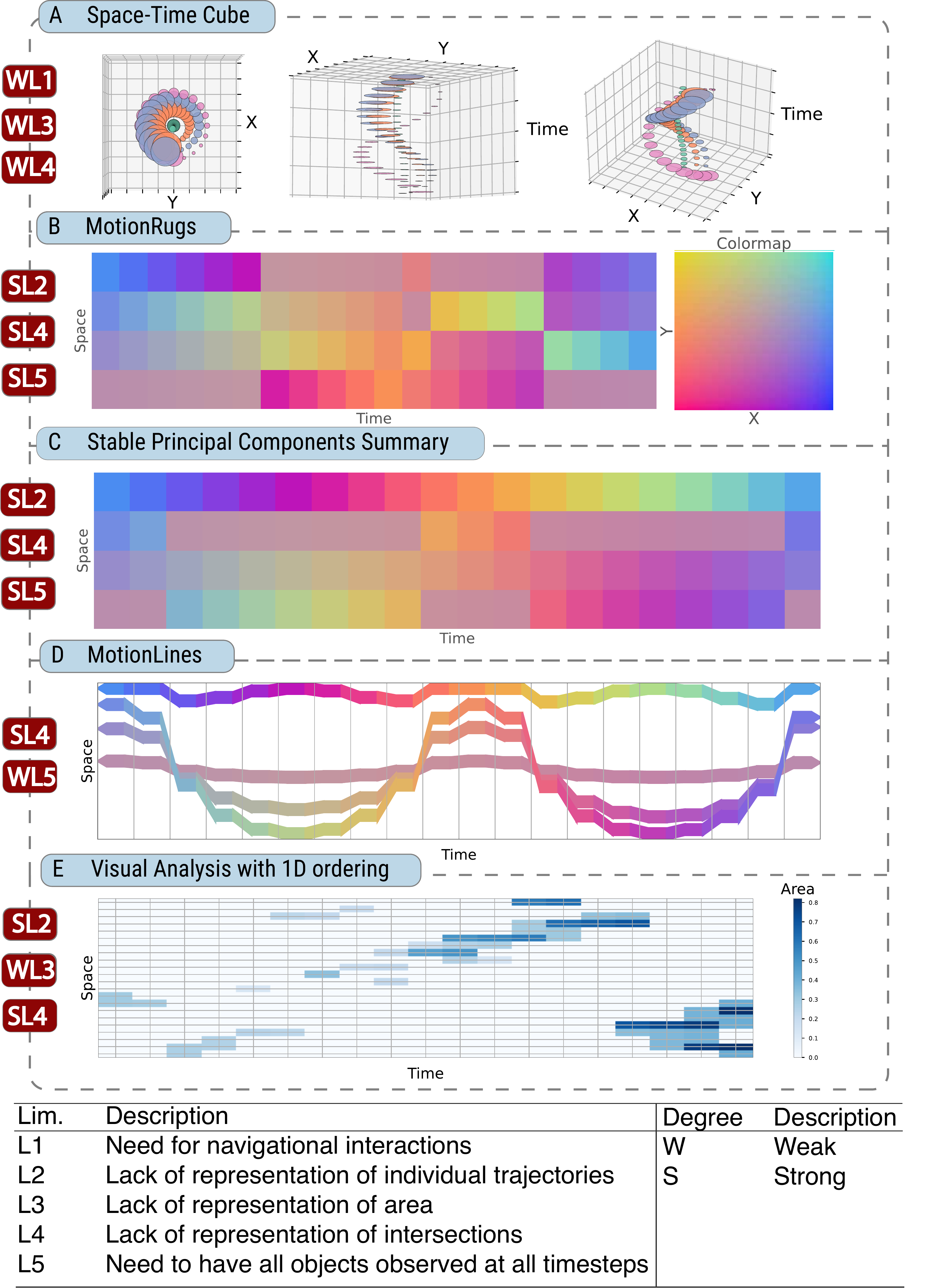}
    \vspace{-0.6cm}
    \caption{
    Results of visualizing a synthetic dataset with different spatiotemporal visualization techniques and identified limitations. 
    We summarize each technique's limitations using the coding scheme in the table above. 
    }
    \vspace{-0.5cm}
    \label{fig:reference_visualizations}
\end{figure}
}

\newcommand{\figPipeline}{
\begin{figure*}[ht!]
    \centering
    \includegraphics[width=\textwidth]{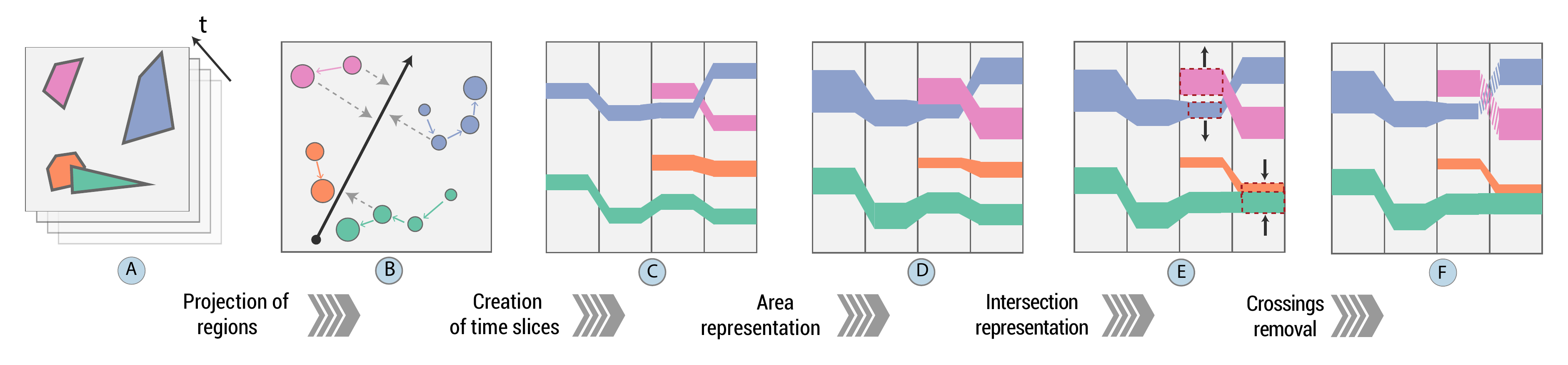}
    \vspace{-0.5cm}
    \caption{An overview of the MoReVis technique. (A) the input data is formed by convex regions that change over time. (B) the regions' from all timesteps (bigger circles indicate later timesteps) are projected to 1D. (C) the time slices are created with a rectangle for each polygon in the corresponding timestep, and the vertical position is the projection value. Rectangles from the same object are connected, forming a curve. (D) the height of the curve represents the area of each region. (E) the vertical positions are adjusted to represent the intersections in the original 2D space. (F) Crossings that do not represent intersections are removed by changing their visual encoding.}
    \vspace{-0.5cm}
    \label{fig:method}
\end{figure*}
}

\newcommand{\figIntersectionLimitation}{
\begin{figure}[t!]
    \centering
    \includegraphics[width=\columnwidth, trim={0 35pt 0 15pt}]{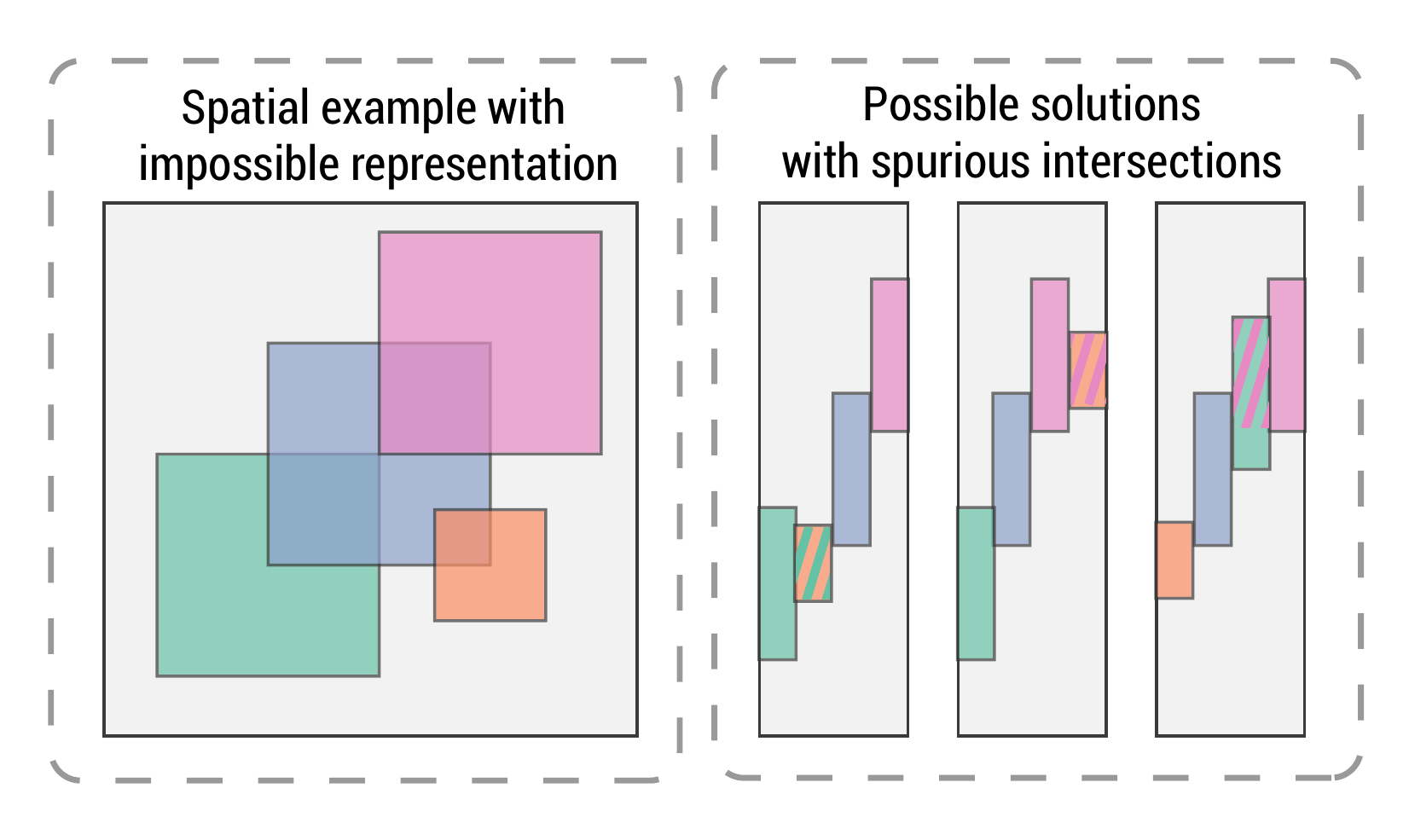}
    \vspace{-0.15cm}
    \caption{This example shows that it might not be possible to accurately represent all the 2D intersections in our 1D scheme, depending on the data. In the 2D space, the purple square intersects three other squares that do not intersect each other, as seen on the left. On the right, we have three different layouts of vertical positioning of the rectangles; the width of the rectangles is reduced to highlight the vertical intersections. In all of the 1D representations, spurious intersections (marked with hatched fills) appear that were not in the original space.}
    \vspace{-0.4cm}
    \label{fig:intersection_limitation}
\end{figure}
}

\newcommand{\figIntersectionAlgorithm}{
\begin{figure}[t!]
    \centering
    \includegraphics[width=\columnwidth, trim={0 30pt 0 0pt}]{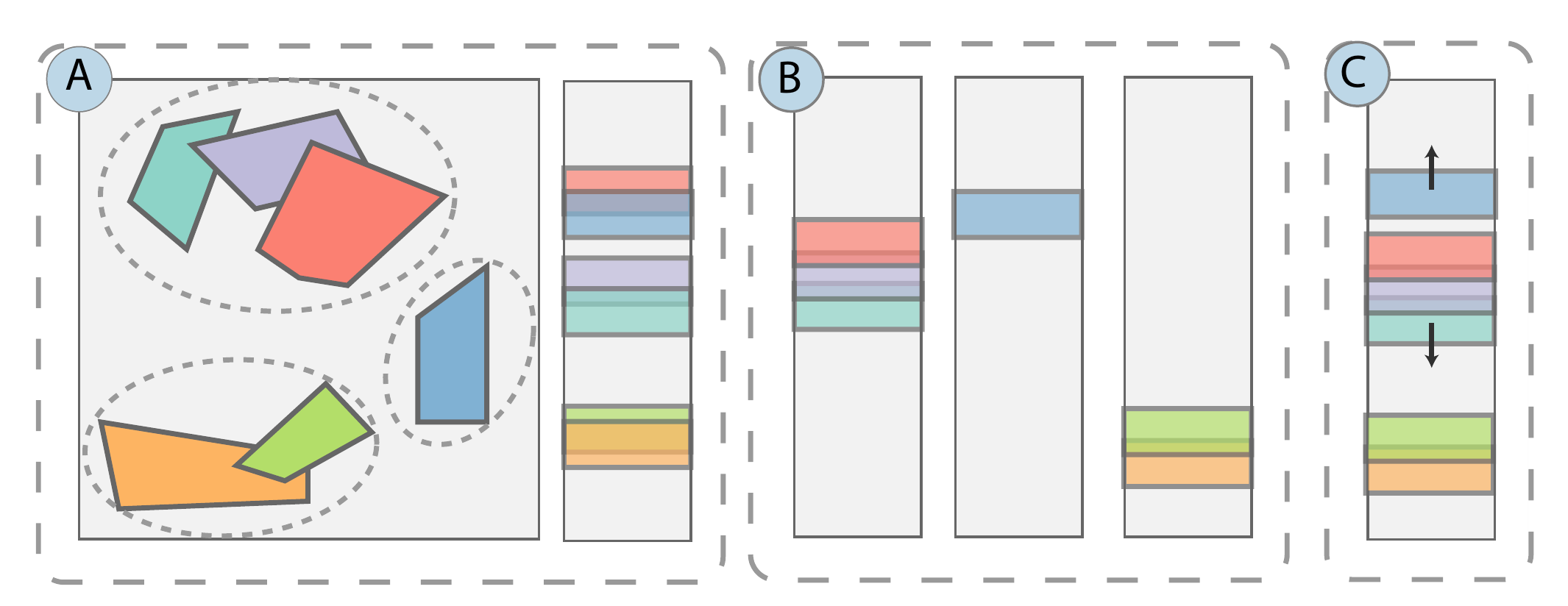}
    \vspace{-0.2cm}
    \caption{(A) Our method first separates the objects into groups that intersect each other in each timestep.
      (B) the intersection representation optimization is used in each group.
      (C) the results of the last phase are merged with another optimization, where the arrows indicate that the groups were shifted to not intersect.}
    \vspace{-0.4cm}
    \label{fig:intersection_algorithm}
\end{figure}
}

\newcommand{\figInterface}{
\begin{figure*}[t!]
    \centering
    \includegraphics[width=\linewidth, trim={0pt 65pt 0 15pt}]{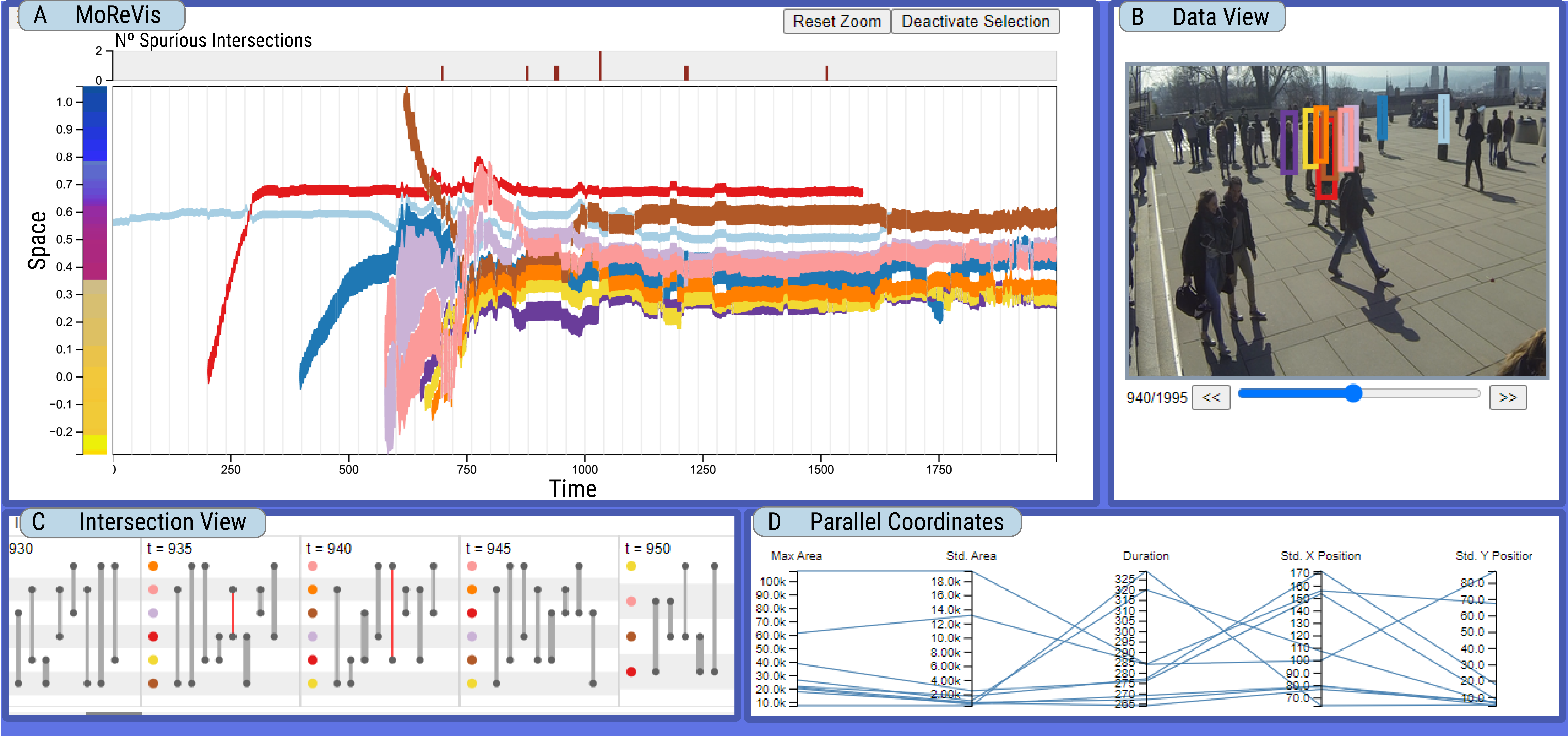}
    \caption{Visual interface implemented used to explore moving regions datasets.
The interface comprises multiple coordinated views': (A) the MoReVis view, (B) the Data View to visualize the original dataset, (C) the Intersection View allows the detailed inspection of the spatial intersections, and (D) the Parallel Coordinates plot that enables the filtering of the curves based on different attributes.}
      \vspace{-0.3cm}
    \label{fig:interface}
\end{figure*}
}

\newcommand{\figProjectionsComparison}{
\begin{figure}[t!]
    \centering
    \includegraphics[width=\columnwidth]{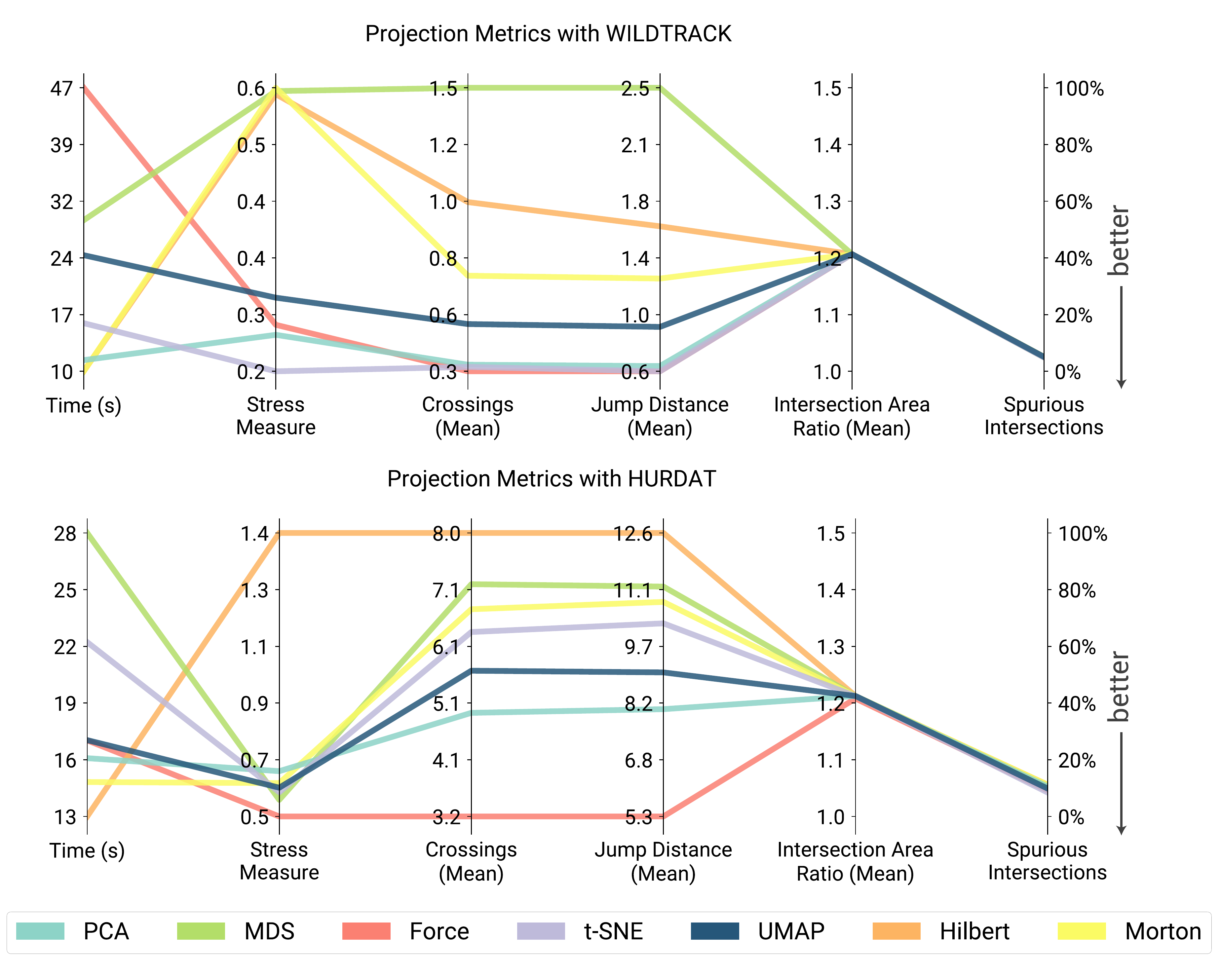}
    \caption{Comparison of all projection techniques (after parameter selection) with the two considered datasets. Each line represents a projection, and each axis is a metric. The different projections present no impact on the two metrics regarding intersections representation; however, considering the other metrics, PCA and force-directed obtained the best results on both datasets.}
    \label{fig:projections_comparison}
\end{figure}
}

\newcommand{\figOptimizationMetrics}{
\begin{figure}[t!]
    \centering
    \includegraphics[width=\columnwidth]{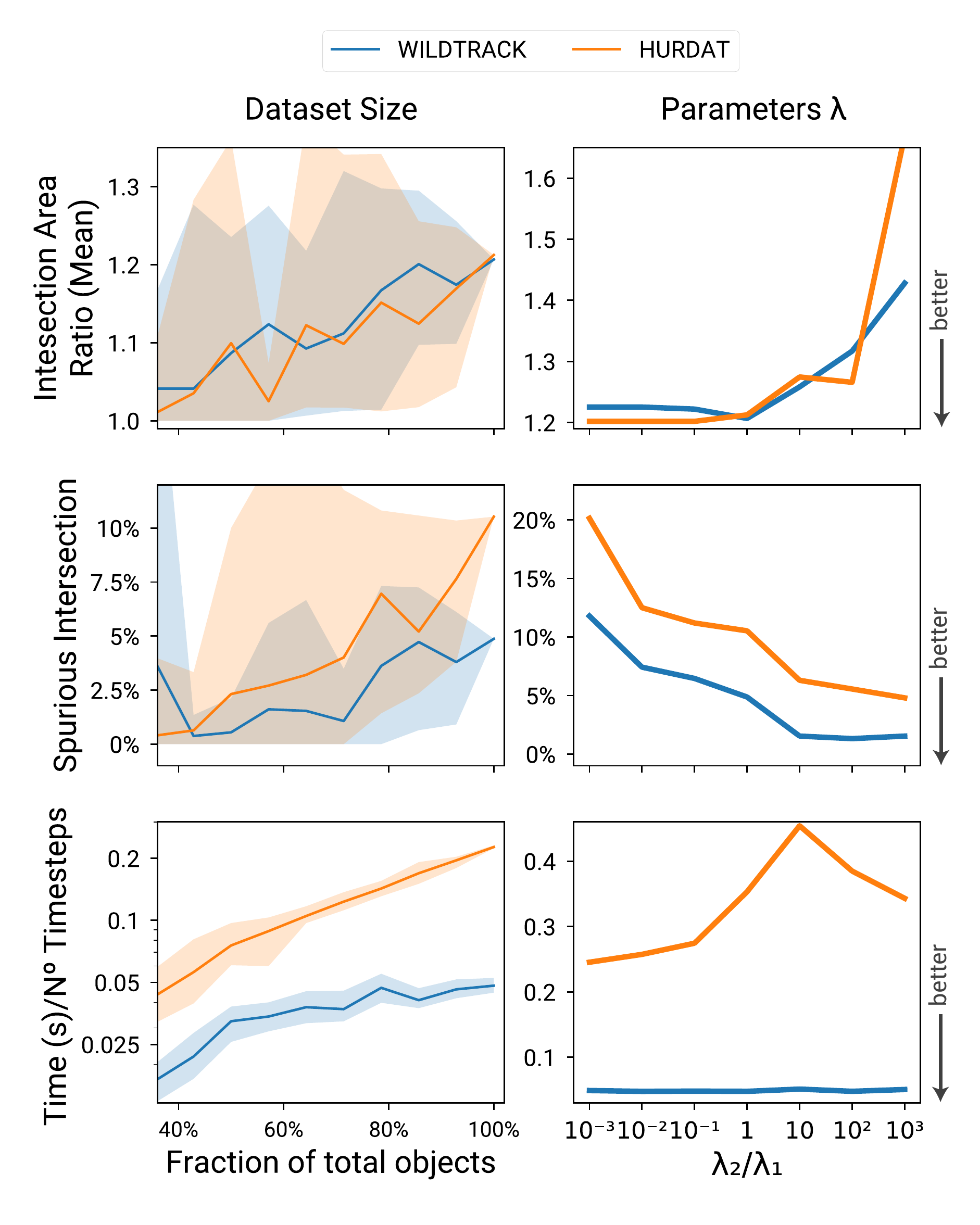}
    \vspace{-0.5cm}
    \caption{Analysis of the optimization results varying the dataset size and the ratio of parameters $\lambda_2/\lambda_1$. The WILDTRACK has at max 14 objects and the HURDAT 70. The optimization method can present lower quality and computing time in bigger datasets. According to the  $\lambda$ parameters values, there is a trade-off between spurious intersections and intersection area ratio.}
    \label{fig:optimization_metrics}
\end{figure}
}

\newcommand{\figMotionlinesComparison}{
\begin{figure*}[t!]
    \centering
    \includegraphics[width=\linewidth]{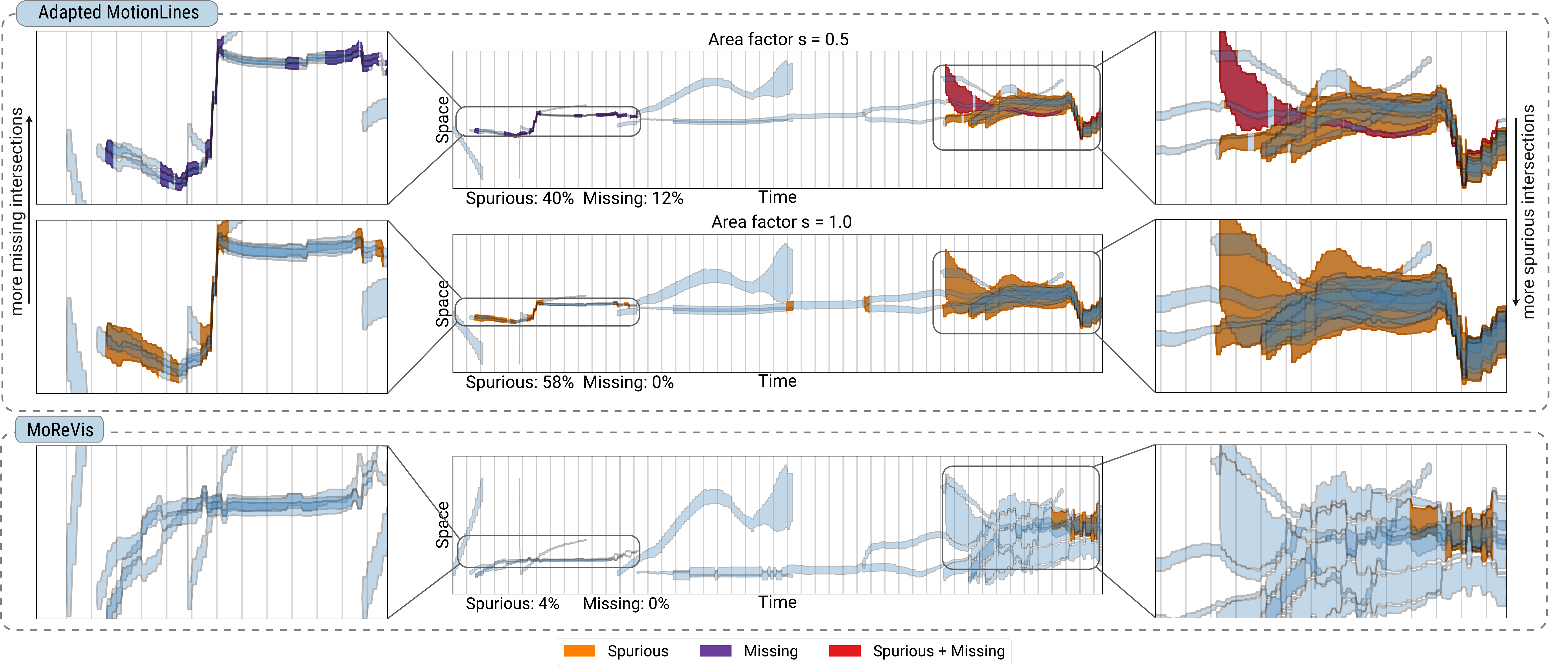}
    \vspace{-0.5cm}
    \caption{Comparison of adapted MotionLines that represent areas and MoReVis. We present the adapted MotionLines result with two area factors to demonstrate the impact of the width of the curves on the intersections. The curves are colored to indicate the presence of errors. There is a trade-off based on the area factor, with $s = 0.5$ there are more missing intersections while with $s = 1.0$ more spurious intersections occur. MoReVis can correctly represent most of these intersections.
    }
    \label{fig:motionlines_comparison}
\end{figure*}
}

\newcommand{\figMotivatingMoReVis}{
\begin{figure}[t!]
    \centering
    \includegraphics[width=\columnwidth]{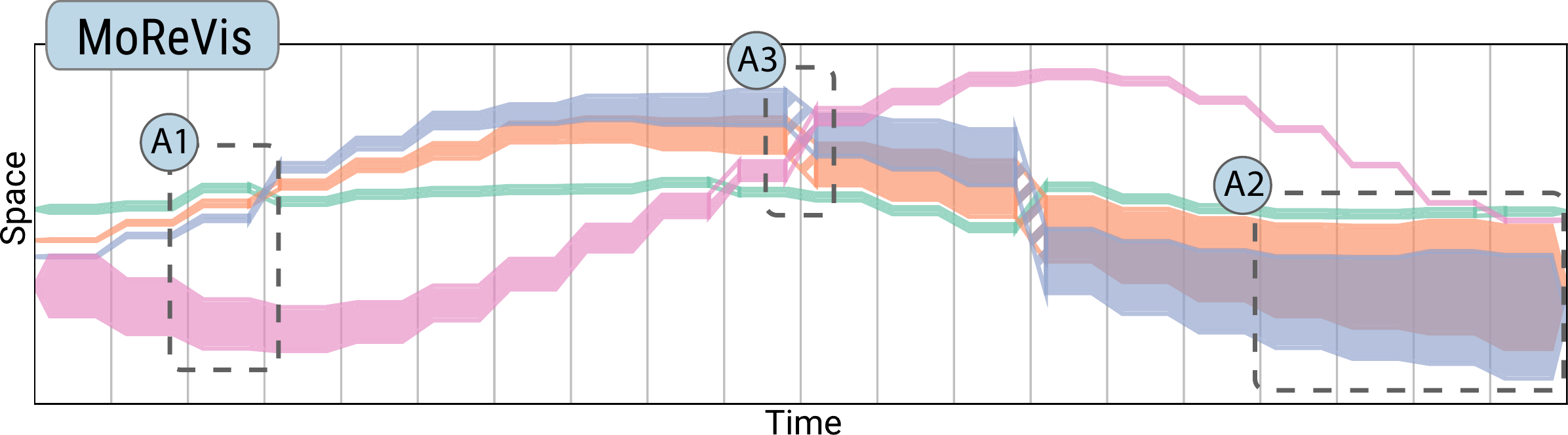}
    \vspace{-0.5cm}
    \caption{
    Result of using MoReVis in the synthetic dataset presented in Fig.~\ref{fig:reference_visualizations}. The technique provides a visual overview suitable for moving regions in which object areas (A1) and spatial interactions (A2),(A3) are taken into account.
    }
    \vspace{-0.4cm}
    \label{fig:motivating_morevis}
\end{figure}
}

\newcommand{\figWildtrack}{
\begin{figure*}[t!]
    \centering
    \includegraphics[width=\linewidth]{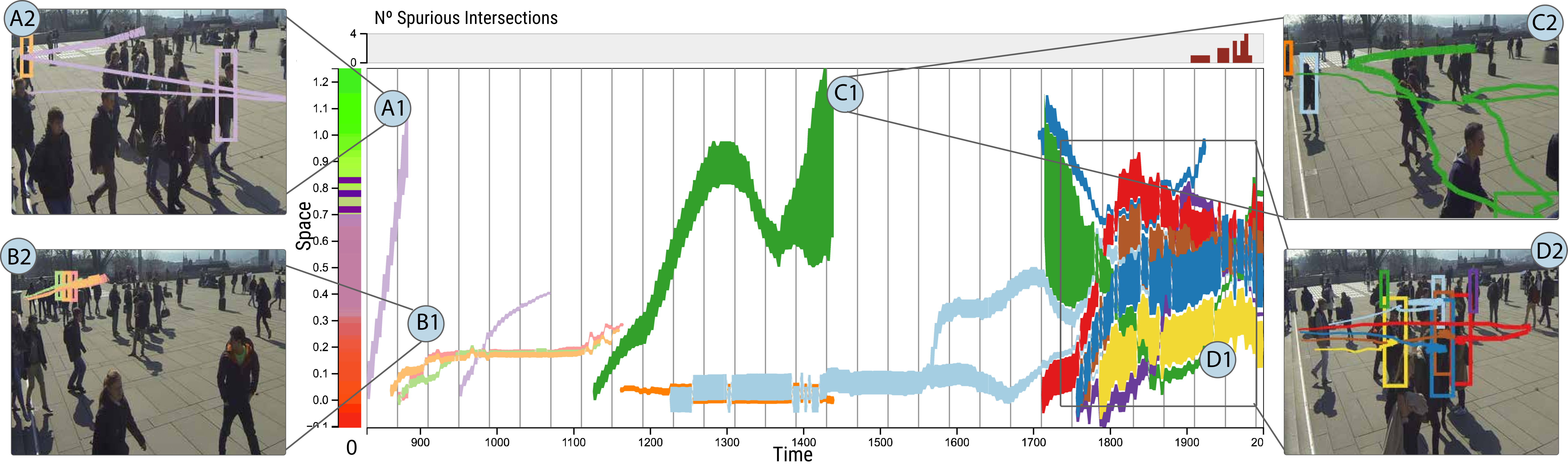}
    \caption{MoReVis result of a selection of the WILDTRACK dataset. Both areas and intersections provide a spatial context to the summary, as shown by the annotated examples. (A1, A2): a pedestrian walks in a zigzag; (B1, B2) a group of three people walks together far from the camera; (C1, C2), a person gets close to the camera, increasing the area of the corresponding bounding box; (D1, D2) a group of friends meets at the center of the screen}
\label{fig:teaser}
\end{figure*}
}

\newcommand{\figHurricane}{
\begin{figure*}[t!]
    \centering
    \includegraphics[width=\textwidth]{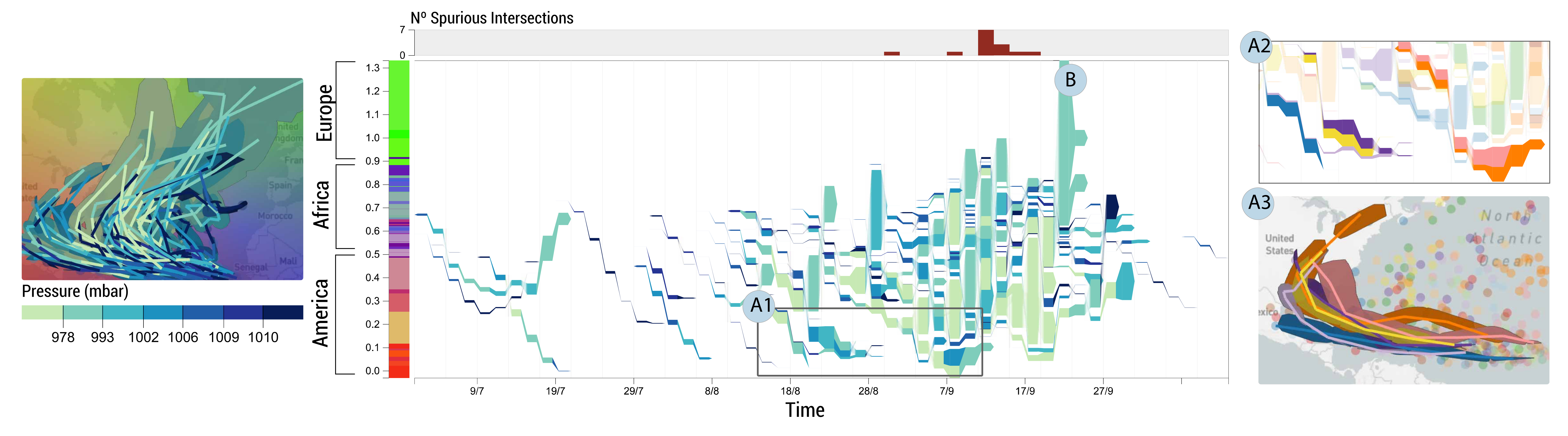}
    \vspace{-0.6cm}
    \caption{
    Usage scenario with a selection of the HURDAT dataset. The y-axis at the left indicates the dominant region of the rectangles in the interval. It is possible to identify a general trend of trajectories that start in Africa, go to America, and return to Europe. On A1 is possible to see two groups of hurricanes that intersect each other in a similar region and separated time intervals, they are highlighted at A2 and A3. 
    On B is marked hurricane Karl which presented one of the biggest areas of action in the observed period.
    }
    \vspace{-0.4cm}
    \label{fig:hurricane-use-case}
\end{figure*}
}

\newcommand{\figUserStudyResults}{
\begin{figure}[t!]
    \centering
    \includegraphics[width=\columnwidth]{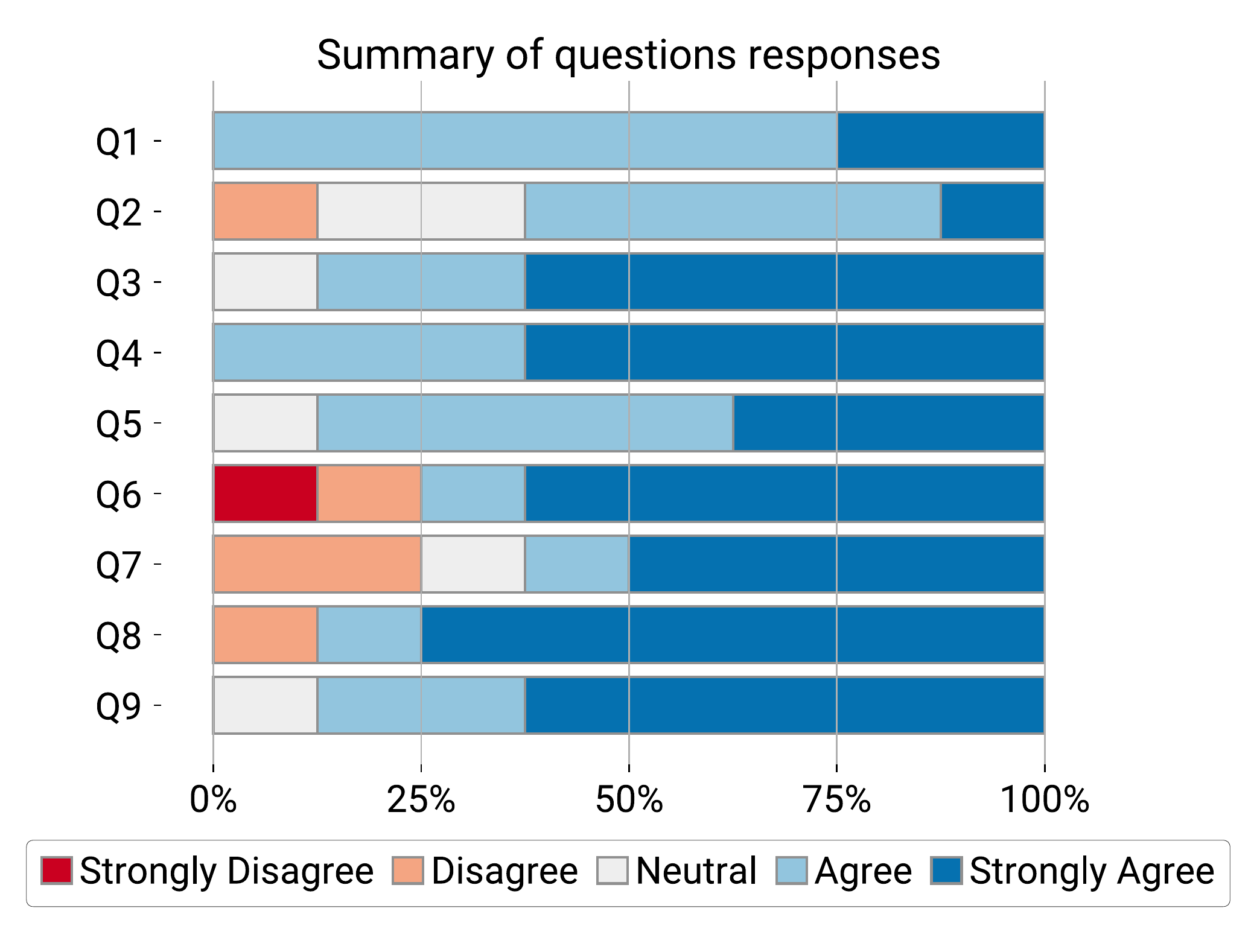}
    \vspace{-0.2cm}
    \caption{Answers for questions Q1-Q9 from 8 participants about the effectiveness of MoReVis. The bar length indicates the percentage of respondents who have chosen that specific Likert level. The predominance of blue tones indicates a positive opinion about the technique.}
    \vspace{-0.4cm}
    \label{fig:user-study-results}
\end{figure}
}

\vgtcinsertpkg

\begin{document}


\maketitle

\section{Introduction}

The wide availability of data acquisition devices has produced large trajectory datasets.
These datasets compile movement data in a span of domains.
For this reason, the construction of analysis and visualization techniques, strategies, and tools to support the exploration of this data type has been a well-studied problem.
Moving entities are often represented as points (dimensionless objects) when studying trajectories. However, in applications such as climate science and video surveillance, the moving entities have extents that are important for analyzing these datasets.
The spatial extent also leads to interaction between objects when there is a spatial intersection. This relationship can contain valuable information about the objects.
We call these types of trajectories representing the movement of objects with a spatial extent as \emph{Moving Regions}.

\looseness=-1
An essential problem in trajectory data visualization is the construction of visual overviews to summarize the movement of a collection of objects in a static plot.
The most straightforward solutions are aggregation, small multiples, or animation-based visualizations.
Aggregation often breaks up trajectories into pieces to form collections that cause the loss of overall movement.
On the other hand, while strategies based on small multiples can give some temporal context, they are limited to a small number of possible timesteps to be shown.
In addition, animations containing many moving objects pose a high cognitive load to the users~\cite{AIGNER:2011:VISUALIZATION,Harrower:2007:Cognitive}. 
Another possible solution is to use a three-dimensional representation, such as the space-time cube metaphor, which uses the two dimensions to represent space and the perpendicular third dimension to represent time.
Nevertheless, this suffers from the usual flaws due to the use of 3D, such as occlusion and perspective distortion.

Recently proposed techniques such as \emph{MotionRugs}~\cite{Buchmuller:2018:MVCTST} (and its variations) attempt to overcome these problems by creating an overview using a 2D metaphor in which the time is represented on one axis and space on another. However, they do not consider the extension of the objects, generating additional problems such as the preservation of intersections in the original space. 
\emph{Storyline} visualizations were initially used to display the narrative of movies in a 2D plot, focusing on the representation of meetings between actors along the movie duration. Most recent works have applied it to more general datasets and different notions of interaction.
While powerful, these other summaries did not consider the combined representation of trajectories, spatial extent, and interactions and are unsuitable for depicting overviews of \textit{moving regions} datasets.

In this paper, we propose MoReVis (\textbf{Mo}ving \textbf{Re}gions \textbf{Vis}ualization). This visualization technique addresses the abovementioned limitations and provides an overview of moving regions. MoReVis uses a 1D representation of space similar to MotionRugs to build an overview as a static 2D plot. We formulate the layout strategy as an optimization problem to properly represent the moving regions' extents and their spatial interactions, in this case, intersections. The final layout ensures that the object's areas and interactions on the visual summary are as close as possible to the areas in the original 2D space.
The final plot illustrates each object as a \emph{curved ribbon}, which uses discrete time in the horizontal direction and space on the vertical axis. We also provide rich interactive features to help users understand the underlying data.
We implement our method in a visual interface and present two usage scenarios using datasets from different domains. These examples show how our approach can provide an overview allowing users to grasp patterns and interactions within the moving region dataset quickly.
Finally, we evaluate MoReVis' effectiveness using numerical experiments and a preliminary user study.
Users were able to answer questions about the dataset under evaluation adequately. In addition, the feedback on our proposal's usefulness and effectiveness was positive overall.
%

In summary, our main contributions are: 
\squishlist
    \item A novel technique for creating a visual summary of moving regions, preserving areas, spatial distances, and intersections between regions as much as possible. 
    \item Visual and interactive tools for better understanding the space transformation utilized and the representation of intersections.
    \item A quantitative and qualitative evaluation of MoReVis, including a comparison with other spatiotemporal visualization methods and a user study.
\squishend

Finally, all the data and code used in this paper are publicly available at \href{http://visualdslab.com/papers/MoReVis/}{http://visualdslab.com/papers/MoReVis/}. 

\section{Related Work}

Our work draws on three streams of prior work trajectory visualization and application domains of moving regions.
Trajectory visualization is a well-studied problem and a complete overview of visual motion analysis; we recommend the survey by Andrienko~et~al.~\cite{Andrienko2013Visual}.
The following two subsections consider trajectories represented as moving points. We discuss classical static summaries of trajectory visualizations and storyline visualization. The last subsection depicts some applications where objects with spatial extent are essential.

\myparagraph{Trajectory Visualization:}
A common problem in visualizing trajectory data is providing an overview of a dataset. 
The most commonly used method for representing trajectories is a static spatial view (often a geographical map), where polylines are used to show the trajectories followed by the moving entities present in the data~\cite{thudt2013visits, wang2011interactive}. 
However, this approach presents a poor representation of the time dimension and can suffer from overplotting. Therefore, variations have been proposed to overcome these problems using aggregation or pattern extraction algorithms that segment the data into consistent motion patterns~\cite{ferreira2013vector,Wang2014Urban}.
On the other hand, the space-time cube~\cite{spacetimecube} uses a different approach to solve these issues by using a 3D-based visual metaphor to represent time as one of the dimensions in 3D space.
This method has been widely used in previous work~\cite{bach2017descriptive,andrienko2013space,chen2015survey}.
However, as it uses a 3D environment, it can present diverse problems: cognitive overload, distortion of distances, and occlusion~\cite{Andrienko2013Visual, ware2013information,evaluating2020filho}.

\noindent More recent methods use static temporal visualizations to avoid the problems present in the space-time cube. These consist of a dense plot with time on the horizontal axis and a discrete set of vertical positions representing the spatial component of the trajectories. As a result, these methods can display the entire period of data and do not suffer from overplotting and occlusions.
These techniques use spatial transformations that preserve the order of positions without considering distances. 
\textit{MotionRugs}~\cite{Buchmuller:2018:MVCTST} is the first representative of this category, proposed to provide an overview of collective motion data. As the objective is to identify the general trend in the motion of a population, it does not represent the individual trajectories and only shows relative distances. 
A further variation of this technique, called \textit{SpatialRugs}, was presented by Buchm\"uller~et~al.~\cite{spatialrugs}, which uses colors to represent absolute spatial positions. 
Subsequently, Franke~et~al.~\cite{1dordering} adapted this idea to present a temporal heatmap to visualize the propagation of natural phenomena. 
JamVis~\cite{2022-JamVis} utilized the 2D representation to show urban events formed by groups of spatiotemporal points.
A vital aspect of these visualizations is how to represent 2D spatial coordinates in one dimension. There are different alternatives to accomplish this task: dimensionality reduction techniques~\cite{Ayesha2020}, spatial indexing methods~\cite{lu1993spatial,GUO:2006:SPATIAL}, or even specially designed projection techniques~\cite{stablevisualsummaries} to improve temporal stability in the results. 
Unlike our work, these techniques focus on collective movements. Thus, they lack a direct representation of individual objects, which is of utmost importance in our case. In addition, and more importantly, they were not designed to represent moving regions (i.e., objects with a spatial extent).

\myparagraph{Storyline Visualizations:}
This group of techniques communicates the evolution of relationships between different objects over time. In general, these relations are the interaction of two objects at the same spatial position.
Commonly referred to as Storyline visualizations, they were often used to represent movie plots. However, lately, they have been used to describe relationships between more generic temporal objects~\cite{pena2021hyperstorylines}. 
In this category of visualizations, entities are represented by curves with a horizontal temporal scale. The vertical proximity between the curves indicates a relationship. 
This group of methods has developed by improving the layout of the curves (reducing line crossings and wiggles)~\cite{van2016block, Tanahashi2012Design, Liu2013Storyflow} or by designing tools that allow the user to control the visualization~\cite{Tang2019istoryline}. 
The closest work related to our proposal in this category was proposed by Arendt~and~Pirrung~\cite{Arendt2017They}, who explicitly incorporated spatial information to create the 1D representation of space. 
Their user study found that explicitly using spatial information improved performance on overview tasks compared to methods that implicitly represent space through object interactions. 
Unlike our proposal, all previous works focus on maintaining the local spatial ordering of objects without preserving the distances\,---\,our work intents to resemble the original distances and intersections between objects as much as possible.

\myparagraph{Moving Regions Applications:}
We now discuss some domains where spatial distances, object areas, and interactions, primarily due to spatial intersections, are vital to interpreting the data. 
The first application is video surveillance systems used in traffic management and monitoring public places, which are essential in intelligent cities~\cite{lisecure2021}.
In this type of video, it is common to perform object detection~\cite{joshi2012survey} automated. 
However, human interaction still plays an essential role in their analysis, using visualizations as support~\cite{raty2010survey}. 
In this context, there are also methods using the space-time cube~\cite{meghdadi2013interactive}, exhibiting the same disadvantages described above. 
On the other hand, Lee and Wittenburg~\cite{lee2019space} use an approach similar to \textit{MotionRugs}~\cite{Buchmuller:2018:MVCTST} but with this type of data. Their method uses the vertical axis to represent time and the horizontal axis to map the horizontal axis of the video frames. 
A limitation in their work is that the projection from 2D to 1D is effortless, creating overlapping plots between objects when there is no spatial interaction. For example, in videos of cars on a highway seen from the front, it is expected that the cars are horizontally aligned, generating many errors. 
Fonseca~and~Paiva~\cite{fonseca2021system} use time bars with interactivity tools to indicate the intervals where the meeting between observed people and other events occurs to enable fast video analysis.

Another recurrent application is trajectories with uncertainty and movement prediction~\cite{domingo2012microaggregation,bonchi2009privacy}. In particular, hurricane trajectories are one theme where the study is necessary for preparation for future events~\cite{Cox2013Visualizing}.
This type of data presents a trajectory formed by recording the positions of the hurricane at different points in time. In addition, these records may include other measurements, such as pressure or wind speed. The use of visualization tools is customary in these data. 
For example, Li~et~al.~\cite{Li2011MoveMine} used spatial mining techniques to decompose hurricane tracks and identify critical features. 
Wang~et~al.~\cite{wang2011interactive} used a map view with trajectories linked to a parallel coordinate plot, a theme river chart, and scatter plots to represent the temporal aspect of other measured attributes. 
With thunderstorm data, Diehl~et~al.~\cite{diehl2021hornero} proposed a tool that used the TITAN algorithm~\cite{dixon1993titan} to obtain regions of the presence of the storms for each timestep and presented a graph abstraction to display the splits and merges along time. 
This modeling of thunderstorms as regions can also be applied to hurricanes. Depending on wind speed and pressure, a hurricane can affect surfaces of different sizes; therefore, it is possible to consider the presence of a hurricane as a region. 

These applications use the same spatiotemporal visualization techniques presented above with some adaptations depending on the domain. Therefore, they have the same limitations, and there is room for improvements in the representation of moving regions. In the next section, we will discuss in detail the shortcomings of some of these techniques.

\figReference

\section{Background and Motivation}
\label{sec:motivating-example}


This section presents an example using a synthetic dataset to introduce our method goals and compare them with related techniques.
This data set consists of four circular objects moving through time in orbits of different radii, as shown in Fig.~\ref{fig:reference_visualizations}(A).
The trajectories of the objects have different behaviors: the green object moves within a small radius; the orange and blue ones move closely, overlapping in the second half of the observed period; finally, the pink object moves in the opposite direction.
In addition, the areas of the objects also have different behaviors: the green object has a constant radius; the pink one has a decreasing radius; the other two have an increasing radius as a linear function of time, leading to an almost complete overlapping of the circles at the last timestep.
This dataset simulates applications such as object tracking in videos, with a bounding box in each frame for each moving object in the scene. This region (bounding box) can change position and shape between different timesteps.

A challenge in this context is providing a visual overview that summarizes the spatiotemporal features of the movement in a given dataset.
Such an overview needs to support the identification (both in space and time) of i) trajectories of individual objects, ii) area changes (which can indicate moving closer to/farther from the camera), and iii) intersections between objects (which can indicate encounters or occlusion between objects).
These tasks were inspired by applications such as analyzing object tracking in videos~\cite{meghdadi2013interactive,hoeferlin2013interactive} and how the StoryLines~\cite{Tanahashi2012Design, Liu2013Storyflow} visualization summarizes the similarities/encounters in a given dataset.

We considered using five previously proposed techniques to provide a visual summary of our synthetic data set (see the results in Fig.~\ref{fig:reference_visualizations}).
In each of the results, we discuss five limitations: 
L1) need for navigational interactions, 
L2) lack of representation of individual trajectories, 
L3) lack of representation of the area, 
L4) lack of spatial overlap (intersections) representation, and 
L5) need to have all objects observed at all timesteps in the time window under consideration.
We consider that techniques can suffer from these limitations in a weak (W) or strong (S) way.


%
\myparagraph{Space-time cube~\cite{spacetimecube}:} 
This technique uses a 3D metaphor in which the horizontal plane represents the spatial positions and the vertical axis the temporal dimension.
We can use the original spatial coordinates to describe the objects' area and intersections directly.
Fig.~\ref{fig:reference_visualizations}(A) illustrates three views of the same spatiotemporal cube.
The first one only shows the spatial information of the objects since it is a view from the top of the cube.
The 3D view has some drawbacks, such as perspective distortion and occlusion, which occur when projecting the cube on a 2D screen~\cite{bach2017descriptive}.
The distortion can hinder the perception of objects' areas (WL3); for example, in Fig.~\ref{fig:reference_visualizations}(A), only the first point of view presents the areas proportional to the actual values.
In each point of view, there are two types of object overlaps: real intersections and visual intersections caused by projecting objects of different depths~\cite{bach2014visualizing} (WL4). 
The user would need to verify which intersections are correct from other points of view.
The 3D-based navigation could be more demanding to control~\cite{evaluating2020filho, sss-gi2001} (WL1), and for that reason, it does not present a fast overview of the data.

\myparagraph{MotionRugs~\cite{Buchmuller:2018:MVCTST}:}
This method is a dense representation where each column represents a timestep, and each cell is a different object. Note that two cells in the same row do not necessarily represent the same object. 
Since this technique was developed for point data, we use the centroid of each region as its position to construct the visual summary in Fig.~\ref{fig:reference_visualizations}(B).
The objects are positioned vertically in each column according to a spatial ordering, in this case, obtained using a Hilbert Curve.
Unlike the space-time cube, the time dimension is represented linearly, and space can be interpreted as the dependent variable. 
We incorporate this representation of time in MoReVis.
The colors in each cell can be based on a feature of the data, such as velocity or area. 
In our example, the colors represent the 2D spatial position of the objects following the 2D color map on the right. 
In this way, it is possible to identify the position of objects in space, not just their relative position, by referring to the 2D color map.
In addition, this coloring also helps to estimate the distances between objects at each timestep. A more significant color change indicates a more considerable distance.
However, this representation has some limitations.
For instance, it is impossible to understand the movement of individual objects (SL2) since this technique independently uses the spatial ordering strategy at each timestep. Thus, the track of each object at different timesteps is lost. 
Furthermore, although it is possible to use colors or glyphs with sizes proportional to the area within the cells, it needs to be clarified how to adapt these metaphors to represent the spatial intersections of the different objects (SL4).
Lastly, this technique needs the same objects to be observed in all timesteps (SL5).

\figMotivatingMoReVis 

\figPipeline

\myparagraph{Stable Principal Components Summary~\cite{stablevisualsummaries}:} 
This method uses the same visual encoding as \textit{MotionRugs}; however, it adopts a modified PCA projection called \emph{Stable Principal Components}~\cite{stablevisualsummaries}. 
This adaptation applies PCA to each timestep; then, it interpolates the results to generate continuous changes in the calculated principal components.
This space transformation method can better represent 1D object trajectories and give a better view of space, as shown in Fig.~\ref{fig:reference_visualizations}(C). 
However, it suffers from the same limitations as MotionRugs, \ie, it is impossible to identify individual trajectories, there is no representation of the spatial intersections, and the objects must be observed in all timesteps (SL2, SL4, SL5).

\myparagraph{MotionLines~\cite{stablevisualsummaries}:} 
This method was presented with \emph{Stable Principal Components}.
The idea is to use the distance between the objects in the spatial representation (\ie, the 1D projection on the y-axis) instead of positioning the objects in each column in their relative order (see Fig.~\ref{fig:reference_visualizations}(D)).
Compared to the previous method, it is possible to identify the movement of individual objects, a representation that we also use in MoReVis. 
Nevertheless, MotionLines does not consider the representation of areas and intersections, which are essential in some applications, such as surveillance videos.
A trivial modification to represent the area would be to change the width of the curves proportionally to the area of the respective objects.
However, as discussed in Sec.\ref{sec:evaluation}, this change can result in many missing and spurious intersections (SL4).
Furthermore, the spatial transformation used does not consider the case where the number of objects is not constant over time (WL5).
The space of each timestep is transformed separately, and when there is one object or none, the 1D space degenerates.

\myparagraph{Visual Analysis with 1D ordering~\cite{1dordering}:} 
This method creates a heatmap to summarize the dataset. 
As in previous work, each column represents a timestep. Each row corresponds to a grid cell of the divided space grid\,---\,\ie, the positions are discretized into a regular grid. 
Next, the mapping of the cells to the vertical position is obtained by dimensionality reduction, in this case, with the Hilbert curve. 
For each cell, the color represents some measure of the objects' density in the corresponding grid cells. 
In Fig.~\ref{fig:reference_visualizations}(E), we use the sum of areas of the objects in that cell. 
Although we can infer global motion trends in this example, this summary uses an aggregation strategy. Therefore, it does not support the study of individual movements (SL2).
It is possible to represent the area of the objects but only the aggregated area (WL3). As a cell has many objects, it could be interpreted as an intersection, but objects are not guaranteed to intersect (SL4).

In summary, the above techniques have different limitations when representing moving regions. 
Therefore, the visualizations produced could be better for summarizing situations where the areas of objects and their spatial interactions (intersections) are relevant. 
Our proposed technique, MoReVis, solves these limitations by a non-trivial combination/extension of ideas from many existing visual summaries. 
Fig.~\ref{fig:motivating_morevis} shows the result of using MoReVis in the synthetic dataset described above (see a detailed discussion in Sec.~\ref{sec:visual_evaluation}).
We highlight the representation of areas (A1), the representation of the intersections between objects (A2), and visual cues to denote the absence of objects' intersections (A3).
%
The following section describes the technique in detail. 


\section{MoReVis}
\label{sec:method}

This section introduces MoReVis, a spatiotemporal visual summary designed to overcome the limitations of previous methods (Sec.~\ref{sec:motivating-example}).
The algorithm to produce this visual summary consists of five steps (illustrated in Fig.~\ref{fig:method}): \emph{Projection of regions}, \emph{Creation of time slices}, \emph{Area representation}, \emph{Intersection representation}, and \emph{Crossings removal}.
We detail each step in the rest of this section.

\subsection{Projection of regions}

The input of our method is a moving regions dataset, \ie, a set of objects $O = \{O_1, O_2, \dots, O_n\}$ that move over time.
At a given timestep, each object is associated with a convex region in the 2D plane (Fig.~\ref{fig:method}(A)). 
Notice that these regions may change over time.
Furthermore, the objects do not need to be observed in all timesteps, different from related techniques.
Finally, each object can have additional associated time-varying attributes representing either numerical or categorical properties. 

\looseness=-1
The first step in our algorithm is obtaining an initial 1D representation of the spatial movement of each object.
This step aims to capture the spatial context of the dataset.
Projection of punctual data is a prevalent task; however, there is important information on the extent of regions in our situation.
We only considered the region's centroids with projection methods that only support point data; otherwise, the distance between regions was used.

\looseness=-1
The considered projection methods include dimensionality reduction techniques: PCA~\cite{pearson1901liii}, MDS~\cite{kruskal1964multidimensional}, force-directed~layout~\cite{improved:2003:tejada}, t-SNE~\cite{van2008visualizing}, UMAP~\cite{mcinnes2020umap}, and space-filling techniques: Hilbert and Morton curves~\cite{lu1993spatial}. 
These projection methods are data-driven, and we used as input the data of all objects to fit them, ignoring the time information.
In recent visual~summaries~\cite{Buchmuller:2018:MVCTST, stablevisualsummaries}, the projection methods, such as Stable Principal Components, were fitted with the data of each timestep separately or using the data of the current and previous timesteps.
Although this procedure presented positive results, we have a different scenario in which the number of objects may vary over time.
Therefore, the spatial representation can be compromised by applying projections in each timestep separately, resulting in a degraded spatial representation when only one object is observed in a given timestep. 
For that reason, we fit the projection with all points to obtain a general space representation.

\subsection{Creation of time slices}

The MoReVis visualization comprises columns indicating the (discrete) set of observed timesteps. 
We represent an object as a rectangle in each column corresponding to timesteps where the object is observed.
The rectangles are vertically positioned so that their center corresponds to the projection value obtained in the previous step.
In addition, all the rectangles have the same width (60\% of the number of pixels corresponding to a column in the plot) and the same height (which will be adjusted later). 
We connect the rectangles corresponding to the same object to form ribbons representing each object's movement. 
The color of each ribbon can be used to convey either object attributes (\eg, uncertainty), movement attributes (\eg, speed), or identifiers used to distinguish the objects.
In the example depicted in Fig.~\ref{fig:method}(C), colors are used to identify different objects. 

\subsection{Area representation}
\label{sec:area-representation}

The remaining steps of the MoReVis technique aim to adjust the initial layout described so far to represent areas of objects and their intersections.
To help describe these steps, we first set up some notation.

The area of a given object $O_i$ at timestep $t$ is denoted by $a_{i, t}$.
$R_{i, t}$ denotes the MoReVis rectangle associated with this object and timestep. 
The vertical position of this rectangle's center is denoted by $y'_{i, t}$ (value obtained in the projection step) and its height by $h_{i, t}$.
Consider that the positions $y'_{i,t}$ were normalized to the interval $[0, 1]$, so the scale is not dependent on the projection method.

This step aims to scale rectangles' height so that their area (in the MoReVis plot) is proportional to their area in the original 2D space.
To do so, we want a scaling factor based on the objects' overall spatial extent.
Furthermore, this scaling factor has to be the same for all timesteps so that the rectangles' heights are comparable through time.
To do so, we first define $A_t = \sum_{i} a_{i, t}$ as the sum of the area of objects in each timestep and $A_M = \max_t \{A_t\}$.
We then set the rectangle's height $h_{i, t} = \dfrac{a_{i, t}}{A_M}$. 
The intuition behind it is that the sum of the heights for all the rectangles should be less than or equal to 1, and it is exactly only when for the timesteps where the total area occupied by the objects corresponds to $A_M$, and the objects are disjoint.

\figIntersectionLimitation

\subsection{Intersection representation}
\label{sec:intersection_representation}

This final step aims to represent intersections between the objects, \ie, make the rectangles in the MoReVis plot intersect (by changing their vertical position) with an intersection area proportional to the actual intersection in the original 2D space.
This problem is challenging since the intersection patterns in 2D space can be complex (Fig.~\ref{fig:intersection_limitation}).
For this reason, we formulate this as an optimization problem that will try to preserve the given spatial configuration as much as possible.
An optimization problem is going to be formulated for each timestep independently.
Thus, for clarity, the description below will omit the subscript $t$ for all the variables. We now set some notation.
Given a pair of objects $(O_{i}, O_{j})$, $w_{i, j}$ denotes the area of the intersection of their regions (already divided by $A_M$), and $I_{i, j}$ denotes the vertical intersection of their associated rectangles $(R_{i}, R_{j})$.

As shown in Fig.~\ref{fig:intersection_limitation}, it is not always possible to represent all intersections correctly. 
Complex 2D intersection patterns can force the creation of {\emph spurious intersections} on the 1D representation.
For this reason, we set the goals of our optimization as follows:

\squishlist
    \item[(G1)] For every intersection in the 2D space ($w_{i, j} > 0$), we want that the corresponding pair of rectangles also intersect in the 1D space, with the 1D intersection being at least as big as the 2D intersection ($w_{i, j} \leq I_{i, j}$).
    \item[(G2)] We also want the 1D intersections not much bigger than the 2D intersections.
    \item[(G3)] If there is no intersection in the 2D space ($w_{i, j} = 0$), we want to avoid, as much as possible, having \emph{spurious intersections} in 1D (which happens when $I_{i, j} > 0$).
    \item[(G4)] We want to keep the rectangles as close as possible to their original positions obtained in the projection step to keep the space representation.
\squishend

To formulate the optimization problem, notice that the vertical intersection $I_{i, j}$ between two rectangles is a function of their height and vertical position.
With the height fixed, if there is no intersection $I_{i, j} = 0$, if one rectangle contains the other $I_{i, j} = \min(h_i, h_j)$ and otherwise $ I_{i, j} = \frac{h_i + h_j}{2} - |y_i - y_j|$.

We first separate the pairs of objects in two disjoints subsets, $A = \{(i,j)| w_{i,j} > 0 \}$ and $B = \{(i,j)| w_{i,j} = 0 \}$, \ie, $A$ is the subset of pair of objects that intersect and $B$ is the pair of objects that do not intersect.
For each pair $(i,j) \in A$, we define a constraint in our optimization to achieve (G1) as:

\begin{equation}
  w_{i,j} \leq  I_{i,j} \Leftrightarrow |y_i - y_j| \leq \frac{h_i + h_j}{2} - w_{i, j}    
\end{equation}

Note that we make $G1$ a constraint; for every intersection in the 2D space, we will have the guarantee that it will also be present in the 1D plot (no missing intersection). 

\figIntersectionAlgorithm

\figInterface

\noindent Similarly, to achieve (G2), for each pair $(i,j) \in A$, we define a real optimization variable $k_{i, j} \geq 1$ and add the following constraint:

\begin{equation}
I_{i, j} \leq k_{i, j}w_{i,j} \Leftrightarrow |y_i - y_j| \geq \dfrac{h_i + h_j}{2} - k_{i,j}w_{i, j}  
\end{equation}

Since (G2) states that $I_{i, j}$ should not be much bigger than $w_{i, j}$, we want each $k_{i, j}$ to be as small as possible.
To this end, we define our first loss as:  $F_1 = \dfrac{1}{|A|} \sum_{(i, j) \in A} k_{i, j}$.

For the third goal (G3), we want to minimize the number of spurious intersections. Therefore, we want to obtain $I_{i,j} = 0$, for every pair $(i,j) \in B$.
For this to happen, we must have $|y_i - y_j| \geq \dfrac{h_i + h_j}{2}$.
To count the number of spurious intersections for each pair in $B$, we add a binary variable $c_{i, j}$ and a new constraint of the form:

\begin{equation}
|y_i - y_j| \geq (1 - c_{i, j})\dfrac{h_i + h_j}{2}  
\end{equation}

When $c_{i, j} = 1$, the constraint is redundant; when $c_{i, j} = 0$, there is no intersection between the rectangles. We therefore define our second loss as $F_2 = \dfrac{1}{|B|}\sum_{(i, j) \in B} c_{i, j}$.
Finally, to fulfill (G4), we desire that the update positions $y_i$ are close to the 1D space representation $y'_i$ obtained previously (in order to retain the spatial representation), so we add the following quadratic error penalty: $F_3 = \sum_{i=1}^n  (y'_i - y_i)^2$.
We combine the three losses into a single one by defining two parameters, $\lambda_1 > 0$ and $\lambda_2 > 0$, and the final objective function of our minimization problem is given by: $\lambda_1 F_1 + \lambda_2 F_2 + F_3$. 
This formulation results in a mixed-integer quadratic programming problem. This type of problem can be solved with a branch-and-bound approach~\cite{lee2011mixed}.

We notice that we can reduce the size of our optimization problem (in the number of variables and constraints) by
partitioning the set of objects into groups. Each group contains the objects that form a connected region\,---\,\ie, there is a path of intersections connecting all objects in the group.  
The six objects were separated into three groups in Fig.~\ref{fig:intersection_algorithm}(A). 
This separation then divides our optimization problem into smaller problems that can be solved more efficiently.
As two objects from two different groups should not present an intersection, we only place them after optimizing each group so there is no overlap.
To do so, we use a quadratic program.
For each group $g$, we compute its total height $h_g$, \ie, the size of the interval that contains all rectangles $(y_i - h_i/2, y_i + h_i/2)$ of the group and its mean position $\overline y_{g} = h_g/2 + \min_{i \in g} (y_i - h_i/2)$.
With the groups ordered by mean position, consider the consecutive pair $(g, g')$; we add constraints $y_g + h_g/2 \leq y_{g'} - h_{g'}/2$, so they will not intersect.
We use an objective function similar to $F_3$ to minimize the overall displacement, $\sum_{g} (y_g - \overline y_g)^2$, where the optimization variables are only $y_g$.
We then place the individual groups and use the individual subproblems to place the rectangles internally in each group.

\subsection{Crossings removal}
\label{sec:crossings}

Lastly, representing the objects as curves and creating links between rectangles can lead to undesired crossings between curves.
Again, removing all the crossings while preserving the spatial context in a lower dimension is impossible. 
Some previous works~\cite{stablevisualsummaries, Buchmuller:2018:MVCTST} also identified this problem and proposed alternative projections that try to generate more stable orderings.
In Sec.~\ref{sec:evaluation}, we evaluate different projection strategies, including those mentioned.
However, since even with the use of these stable projections, we still can face undesired crossings we 
decide to represent them visually.
A crossing between two links is \textit{spurious} if the objects in the previous and the next timestep present no intersection in the original space.
Similar to the visual cues studied by Bäuerle~et.~al.~\cite{bauerle2022where} to represent missing information, we changed the encoding of every link involved in a \textit{spurious} crossing to exhibit hashed color or gradient in opacity.

\section{Visualization Interface}
\label{sec:interface}

We implemented MoReVis in an interactive interface (shown in Fig.~\ref{fig:interface}) with four coordinated views:
\emph{MoReVis}, \emph{Data View}, the \emph{Intersection View}, and the \emph{Parallel Coordinates}. 
Each view will be explained in detail in the following sub-sections.

\looseness=-1
\myparagraph{MoReVis View:}
This is the main view in our system~(Fig.~\ref{fig:interface}(A)) and aims to present the MoReVis plot alongside additional visual clues that help in the data exploration process.
This view also supports zooming and panning and presents a tooltip showing detailed information about the object when the mouse hovers over a curve.

Next to the y-axis, a color bar is present to interpret the 1D space and to verify the quality of the projection inspired by the coloring from SpatialRugs~\cite{spatialrugs}. 
A 2D color map is placed on the original space, and the coloring of the objects in each timestep is the color of the centroids' position.
However, instead of keeping colors in the curves, we moved all rectangles to the same horizontal position, blending the color of overlapping rectangles.
The 2D color map is displayed in the Data View by hovering over this color bar.
The idea is to inspect the space's most used regions represented by the vertical axis.
Thus, to evaluate the spatial transformation, we look for abrupt changes in color as neighborhoods in the original space have similar colors.

Finally, the bar chart on the top indicates the number of spurious intersections in each timestep (similar to the error plots presented in \cite{stablevisualsummaries}).
For example, in Fig.~\ref{fig:interface}(A), we can see that the maximum number of spurious intersections in a single timestep is 2.
The rectangles that participate in spurious intersections are highlighted by hovering over any of the bars. 

\myparagraph{Data View:}
This view presents the original moving regions dataset, so the visual metaphor depends on the application domain.
In our application, we implemented two options.
The first one is used for data representing geographical trajectories. In this case, the view presents a 2D geographical map with polygonal lines and shapes depicting the moving regions  as in Fig.~\ref{fig:hurricane-use-case} (Sec.~\ref{sec:use-case-hurdat}). 
The second option works for data representing object tracking in videos (as in Fig.~\ref{fig:teaser} and Fig.~\ref{fig:interface}(B)); we show the video frames with the bounding boxes of the tracked objects. 
Finally, when the user hovers the mouse over a curve in the MoReVis view, the data for the corresponding timestep is shown on the Data View.
Similarly, when a user clicks on a curve, the trajectory of the centroids of the object is shown on the Data View.

\myparagraph{Intersection View:}
This view is activated when the user creates a brush on the MoReVis plot to present details of the structure of the intersections.
For each timestep in the horizontal extent of the brush, we create a graph where nodes are objects (contained in the brush region), and an edge is made if two rectangles intersect. 
We use a vertical layout to display the graphs, similar to recent works~\cite{Valdivia2021Analyzing, Elzen2014Dynamic}. Each row is a node in this display, and a line between two rows is an edge.
Notice that we only show nodes with a non-zero degree.
The width of the edges is proportional to the intersection area, and black edges indicate real intersections. In contrast, the red ones represent spurious intersections (which are not present in the 2D original space).  
For example, in Fig.\ref{fig:interface}(C), it is possible to see a spurious intersection at the timestep $940$ between objects red and pink; on the Data View, it is possible to verify that they are close but present no intersection.
This view was designed to facilitate the verification of details in intersections and depicted possible errors.

\myparagraph{Parallel Coordinates:}
We use a parallel coordinates plot~\cite{inselberg1990parallel} to represent object attributes, such as measures of their overall movement, area, and presence on the video, as shown in Fig.~\ref{fig:interface}(D). 
In the usage scenario with hurricanes (Sec.~\ref{sec:use-case-hurdat}), other attributes were also used: their max velocity, wind speed, and pressure.
Brushing the different axes in this plot allows the user to filter the objects shown in the MoReVis view.

\section{Experimental Evaluation}
\label{sec:evaluation}

This section presents a series of experiments to evaluate the MoReVis algorithm in terms of its parameters.

\subsection{Datasets}
\label{sec:datasets}

Our evaluation uses two real datasets from different domains: object tracking in videos and hurricane trajectories. These datasets are described in the following.

\myparagraph{WILDTRACK~\cite{Chavdarova2018Wildtrack}:} consists of an object tracking dataset produced from a video captured in a public open area with an intense movement of people. 
The data contain the original video and the bounding boxes of tracked people for each frame in the video. 
We only considered a subset of 14 people (moving regions) with a long presence on the video, having a total of 234 timesteps.
We decided to filter the people to generate more interpretable results that our users can evaluate through our visualization.

\myparagraph{HURDAT~\cite{hurdat}:} describes hurricane trajectories tracked since 2004. Each trajectory contains several attributes, such as windy velocity, pressure, and the spatial extent of the storm, with measurements in intervals of 6 hours. 
We used each hurricane as a moving region, and each timestep represents two days. The area is the convex hull that includes all region's measurements of the hurricane inside this time interval.
In addition, we disregard the year of occurrence for each timestamp to investigate seasonal patterns. 
Finally, we selected the hurricanes that started with the longitude inside the interval $[-50, -20]$ and latitude inside $[10, 20]$ (west coast of Africa). It resulted in a total of 70 hurricane tractories along 52 timesteps.

\subsection{Quality Metrics}
\label{sec:metrics}

We now describe the metrics used to evaluate the MoReVis layout.

\myparagraph{Stress Measure:} 
This metric
is commonly used to assess the quality of dimensionality reduction techniques,
it is the difference between the distances in the original and the projected spaces. 
More clearly, let $d_{O_i, O_j, t_p, t_q}$ be the distance between the region of objects $O_i$ and $O_j$ at timesteps $t_p$ and $t_q$, and let $\hat d_{O_i, O_j, t_p, t_q}$ denote the distance between the respective rectangles.
We use the length of the smallest segment that links the regions or rectangles as the distance.
We build the pairwise distance matrices for all $d_{O_i, O_j, t_p, t_q}$ called $D$ and for all $\hat d_{O_i, O_j, t_p, t_q}$ called $\hat D$. Consider that they are divided by the maximum value to remove the scale.
The \textit{stress measure} is $\sqrt{||D - \hat D||_F^2 / ||D||_F^2}$, where $||.||_F$ denotes de Frobenius norm of a matrix.

\myparagraph{Crossing and Jump Distance:}
These metrics were present in evaluating space projection techniques for MotionLines~\cite{stablevisualsummaries}.
The crossings (change of curves order between timesteps) can generate misrepresentations of closeness. 
Consider that $S_{i, t}$ is the ranking of the object $O_i$ on timestep $t$ in relation to the vertical positions.
Given pair of consecutive timesteps, we count the pairs of objects $(O_{i}, O_{j})$ such that $S_{i, t} < S_{j, t}$ and $S_{i, t + 1} > S_{j, t+1}$ (their order changes).
The \textit{crossing metric} is the average crossing calculated for all consecutive timesteps.
Similarly, the jump distance is the difference in the ranking of the objects between timesteps; it is $\sum_{i} |S_{i, t} - S_{i, t+1}|$. 
The sum is over every object present in both timesteps.
Similarly, the \textit{jump distance metric} is the average value for all timesteps.

\myparagraph{Intersection Area Ratio Error:}
On our formulation (Sec.~\ref{sec:method}) for every object's intersection, we allow $I_{i,j}$ to be bigger than $w_{i, j}$. 
Therefore, to evaluate the size of this ``distortion'', we define the intersection area ratio error as the mean of the ratios $I_{i,j}/ w_{i,j}$ for all intersecting pairs of objects ($w_{i, j} > 0$).

\looseness=-1
\myparagraph{Spurious Intersection Error:}
Although there is no missing intersection in the MoReVis plot, spurious intersections may occur.
Therefore we define the \textit{spurious intersection error} as the percentage of intersections represented in a given plot that is spurious.

\figProjectionsComparison

\subsection{Quantitative Evaluation}

\myparagraph{Projection Evaluation:}
Similar to related techniques (Sec.~\ref{sec:motivating-example}), the MoReVis result depends on the method used to transform the original 2D space into 1D.
We evaluate a set of projections with both datasets to identify the most suited projection technique.
We proceed with this evaluation in two steps. 
First, for each projection, we identify the best parameter values.
We used the practices discussed in the literature and performed a grid search with all combinations to identify the best parameter values. 
The details are available in the accompanying supplementary material.
We highlight that we used the region's centroid as the representative position for techniques that directly use positions (\ie, PCA, Hilbert Curve, and Morton Curve). 
On the other hand, when using techniques based on distances (\ie, force-directed layout, MDS, T-SNE, and UMAp), we evaluated the use of distances between their centroids and the distances between the regions (considering the extent).
With our datasets, the cost of computing distances between regions was insignificant and can be more informative considering the extent.
To select the best set of parameters, we chose the ones that minimized the stress measure, crossings, and jump distance with this respective order of priority.

In the second step, we compare the different techniques.
Fig.~\ref{fig:projections_comparison} shows the results, where each line represents a technique, and the vertical axes include the quality metrics (discussed in Sec.~\ref{sec:metrics}) and the overall computing time of the MoReVis technique with each projection technique.
Notice that the choice of the technique has great importance for the time necessary to compute the complete MoReVis result; slower methods can almost double the required time. 
The projection technique had a minor influence on the metrics related to the optimization step in MoReVis: intersection area ratio and the number of spurious intersections.
We can also see that the space-filling curves and MDS did not obtain the best result in any of the metrics.
Finally, while t-SNE and UMAP attained good results, they demand the evaluation of a large number of parameters and involve a high computing time.
Overall, the two best techniques are the force-directed layout and PCA, with similar 1D representations. A different aspect is that PCA considers only the centroids of points, while the force-directed considers the extent. 
One drawback of the force-directed is the iterative process, making it slower than PCA
The good results of PCA can be interpreted because the data is low-dimensional (2D). 
Considering these results, we used PCA in the next steps of evaluations and usage scenarios.

\figOptimizationMetrics

\myparagraph{Optimization Evaluation:} 
Presented in Fig.~\ref{fig:projections_comparison}, the spurious intersection error for WILDTRACK was $5\%$, for HURDAT was $10.5\%$, and the intersection area ratio for both was $1.2$.
These results show that our optimization obtained good approximations despite not getting a perfect presentation of intersections in 1D.

The size of the problem impacts these metrics: the bigger the number of objects in each group (as described in Sec.~\ref{sec:intersection_representation}), the more complex the intersection structure to be represented.
To investigate this dependency, we ran the MoReVis algorithm on a reduced set of objects randomly sampled from the original dataset for each dataset.
More clearly, we generated ten random samples for different sample sizes, which were given as input to the MoReVis algorithm.
The average error measures are reported in Fig.~\ref{fig:optimization_metrics}(left); the filled region marks the min and max values obtained.
Notice that the algorithm can produce layouts with small values of spurious intersection and intersection area ratio for a small set of objects. However, the errors increase as the number of objects grows, and more complex spatial structures happen.
The number of objects also greatly impacts the computation time, as the number of constraints and integer variables grows quadratically with the number of objects. 
Fig.~\ref{fig:optimization_metrics} (bottom-left) shows the total computation time divided by the number of timesteps (the WILDTRACK presents more timesteps but fewer objects than HURDAT). 
The mean time per timestep for small samples (in both datasets) is at most $0.1$ seconds. Still, in the case of large samples for the HURDAT dataset, the computation can take, on average, close to $0.2$ seconds, which results in a total computation time of around 30 seconds.

We now investigate the dependency of the quality metrics on the parameters $\lambda_1$ and $\lambda_2$.
To compare the different situations, we considered different values for the ratio $\lambda_2/\lambda_1$.
When this value is small, more weight is given to minimizing the intersection area ratio. On the other hand, when the ratio is large, more weight is given to reducing the number of spurious intersections. 
The results of this experiment are shown in Fig.~\ref{fig:optimization_metrics} (right). 
The expected relation between $\lambda_2/\lambda_1$ and the metrics is observed.
It is also possible to identify that the HURDAT presented a longer computing time when there was more importance on preserving spurious intersections.
Depending on the application, one can set the parameters to prioritize one of the aspects.
In all the usage scenarios this paper presents, we set both $\lambda_1 = \lambda_2 = 1$.

\figMotionlinesComparison

\subsection{Visual Evaluation}
\label{sec:visual_evaluation}
We now present and discuss the visual results of MoReVis with some of our datasets.
Fig.~\ref{fig:motivating_morevis} shows the technique for the synthetic dataset described in Sec.~\ref{sec:motivating-example}. 
We can identify the scale of the area (A1), with the pink object having a greater extent that is decreasing through time. In contrast, the objects represented in orange and red have increasing regions. 
Furthermore, it is possible to identify that the objects started intersecting close to the middle of the observation period and almost overlapped entirely by the end (A2).
In (A3), we demonstrate the importance of the crossing removal step.
Notice that at this timestep, the pink curve crosses the other two curves that do not intersect in the original space.
The visual cue present in MoReVis indicates this inconsistency, and the final result presents no missing or spurious intersections.
This example illustrates the effectiveness of MoReVis in supporting the tasks discussed in Sec.~\ref{sec:motivating-example}.
The example presented in Fig.~\ref{fig:motionlines_comparison} aims to show the importance of the optimization step in MoReVis.
To do this, we compare MoReVis against an adapted version of MotionLines~\cite{stablevisualsummaries} (using the SPC projection) that represents areas on the WILDTRACK dataset. 
Considering the same notation of Sec.~\ref{sec:area-representation}, in the MotionLines adaptation we set the width of the curves by $h_{i, t} = s\dfrac{a_{i, t}}{A_M}$. This $s$ parameter permits us to generate different scales of the curves of the plot, and we evaluated $s = 0.5$ and $s = 1$.
The curves are plotted in blue, and some rectangles are highlighted by filling in orange, purple, or red if they had a spurious intersection, missing intersection, or both, respectively. 
Notice that this trivial adaptation of MotionLines presents some uncontrolled flaws in the representation of intersections.
There are missing intersections when the curves are thin, as shown on the top plot. When the curves are thick, many spurious intersections appear. This is evident at the end of the observed period, where many close objects have spurious intersections for an extended period.
In the bottom plot, the MoReVis result can adjust the position of curves so that there are no missing intersections and just a few spurious intersections.

In summary, when compared with other techniques, MoReVis can represent objects' area and spatial interactions.
These issues were not considered in previous works.
Furthermore, our results show that while it is true that we can adapt previous techniques to represent the area (\eg, MotionRugs, MotionLines), these solutions are insufficient. 
Therefore, non-trivial solutions are needed to achieve the objective and our proposal points in that direction.
In particular, when considering the more widely used solutions like space-time cubes and animations, MoReVis can provide an overview of all the spatial movement, 
area, and intersections simultaneously without requiring extensive navigational interactions or spatial distortions caused by 3D views.
All of this allows for fast identification of spatial events, temporal patterns concerning spatial extents (areas), and direct comparison across different timesteps.
\section{Usage Scenarios}
\label{sec:uses}
\label{sec:use-case}

We present two usage scenarios to demonstrate the applicability of MoReVis in different domains. 
The visual interface presented in Sec.~\ref{sec:interface} was used to produce the analysis.

\subsection{Object tracking data}
\label{sec:use-case-wildtrack}
\looseness=-1
First, we explore the WIDLTRACK dataset described in Sec.~\ref{sec:datasets}.
It consists of people tracked on a video captured at an entrance of a building.
In the video, many pedestrians pass by that mostly enter and leave the scene quickly.
We applied filtering to select objects that stayed at least $200$ and less than $1000$ frames on the screen. 
The MoReVis result with the PCA projection is depicted in Fig.~\ref{fig:teaser}.
The visualization elucidated several properties of the data, discussed in the following.
At the start, the violet curve on the left (A1) presents a significant change in the vertical position, suggesting a broad movement that passes through a large extent of the scene.
The curve also presents a small width, indicating that the pedestrian has a relatively small bounding box, \ie, is far from the camera. 
Around the timestep $900$, this curve jumps from the top of the plot to the bottom; this was identified as an error in the original data: two persons were classified as the same.
Fig.~\ref{fig:teaser}(A2) shows the trajectory of the original video, forming a zig-zag due to this classification error.
Another example is highlighted as (B1), which consists of three thin curves (light green, orange, and pink) overlapping. With frame (B2), we identify three classified pedestrians walking together with intersecting bounding boxes represented in the plot. As they are distant from the camera, their curves are thin.

The green curve (C1) presents a significant increase in its width. This occurs due to a pedestrian walking in the camera's direction, getting close to it, as shown in (C2).
This pedestrian leaves the screen for a while and later returns (at timestep $1700$) to the viewport.
The curves inside the region (D1) present the most intersections in the observed period. This result relates to a group meeting in the middle of the scene, as depicted in (D2). 
One detail is that the yellow curve does not intersect with the main group (blue, red, and brown). This can be verified in the plot and also in the frame view. However, it crosses with the green object between timesteps $1800$ and $1900$.
In this portion of the visualization, there are also some spurious intersections due to the complexity of intersections of the data. 
The bar chart on the top shows four spurious intersections that can be detailed with the Intersection View.

\figWildtrack

\figHurricane

\subsection{Hurricane data}
\label{sec:use-case-hurdat}

\looseness=-1
A common analysis with hurricane data is comparing the trajectories with close start positions~\cite{camargo2007cluster,tripathi2016direction,mirzargar2014curve}.
In this use case, we perform such a task using the portion of the HURDAT dataset (Sec.~\ref{sec:datasets}) that corresponds to hurricanes that started with longitude between $-50^{\circ}$ and $-20^{\circ}$
and latitude between $0^{\circ}$ and $20^{\circ}$, a region close to the west coast of Africa.
The MoReVis plot generated is presented in Fig.~\ref{fig:hurricane-use-case}.
To interpret the space transformation made by PCA, we annotate the color bar on the left, indicating the dominant regions in different vertical intervals of the 1D space representation. 
The color map is also presented on the map at the left.
We can see general movement trends, with many hurricanes starting close to the west coast of Africa and going towards.
Some more prolonged hurricanes move after towards Europe.
The color on the curves shows the air pressure for each hurricane.
With the coloring, it is possible to identify an inverse relationship between the area and the pressure.
Additional interesting general patterns include that hurricanes usually start with small widths and increase through time.
As MoReVis does not suffer from occlusion and tries to minimize spurious intersections, the hurricanes' trajectories are more easily identified.
This example also shows the importance of the crossings removal step (Sec.~\ref{sec:crossings}).
The intersections in this example are situations where hurricanes pass through the same region at the same period of the year.
At the bottom of the plot, we can identify the example marked with A1. These same hurricanes are highlighted in A2 and shown on a map in A3, with a color update to distinguish them.
We have two separate groups that have intersections for some timesteps, and both occur near the east coast of North America.

\looseness=-1
One hurricane that stands out is marked with (B); it has the most significant area observed and entered the most inside Europe. 
This was hurricane Karl from 2004 which peaked as a category four on the Saffir-Simpson scale.
Its intensity quickly diminished, causing no fatalities when it reached Norway at the end of its trajectory.

\section{User Study}

We performed a user study to evaluate the ease of using MoReVis on analytical scenarios. In this section, we describe the details of the survey and the results.

\subsection{Study Procedure}
We had nine participants in the study (eight undergraduate students and one master's student). Six already had previous experience with spatiotemporal data, and four were familiar with visualization techniques for this type of data. Participation was voluntary and unpaid.
The experiment was conducted asynchronously, \ie, participants were free to run the study at any time.
First, participants had to watch a 10-minute video with a general explanation of spatiotemporal data, a simple explanation of MoReVis, and a description of the visualization web interface. 

Subsequently, the participants had to perform three groups of activities between tasks and questions. The first group was designed to validate if they understood MoReVis. The second group dealt with spatiotemporal aspects of visualization. Finally, the third group was about object intersections.

\myparagraph{Tasks:}
The tasks of group one ($T1$) were: counting the objects in the frame, counting the number of intersections of an object, and identifying the time interval with the most intersections.
The tasks of group two ($T2$) were: identifying the objects with the most movement on the screen, relating regions of the original space and intervals in the 1D space representation, identifying the region of the original space with the largest number of objects, identifying the object with the largest area, and interpreting the event that caused this largest area.
Finally, the tasks of group three ($T3$) were: counting the number of intersections of an object, interpreting these intersections, identifying the object that intersects with another specific object, identifying the time step interval with the largest number of intersections, interpreting the event that caused this largest number of intersections, identifying a pair of objects with spurious intersections in a specific timestep, and identifying all objects with spurious intersections in a specific time step interval.

\looseness=-1
\myparagraph{Questions:}
The questions for tasks $T2$ were: 
(Q1) \emph{``Is it easy to interpret distances on the MoReVis view?"}, 
(Q2) \emph{``Is it easy to interpret the 1D representation of space in the MoReVis view?"}, 
(Q3) \emph{``Does the color bar on the left assist in interpreting the 1D representation of the space in the MoReVis view?"} and 
(Q4) \emph{``Is it easy to identify the area of objects in the MoReVis visualization?"}. 
Using the same format, the questions asked after tasks $T3$ were:
(Q5) \emph{``Is it easy to identify intersections between objects?"}, 
(Q6) ``Is it easy to identify if an intersection is spurious?", 
(Q7) \emph{``Is it easy to identify the region that has spurious intersections?"}, 
(Q8) \emph{``Does the selection in the MoReVis view help to analyze intersections?"}, and
(Q9) \emph{``Does the representation of intersections in the MoReVis view facilitate the data's spatial interpretation?"}. 
We asked users to comment on answers Q2, Q5, Q6, and Q9, asked them which of the three other interface views they found useful and a final comment.

\subsection{Results}

One of the nine participants responded that he did not understand MoReVis; therefore, we eliminated his/her responses from the analysis. As for the other responses to the validation stage, all participants answered the first question correctly. For the other two questions, we had one incorrect answer for each (different participants). With that, we kept eight participants in the analysis of the results. 
When asked about the object with the most movement on the screen in one of the tasks of $T1$, three participants responded with objects with the longest presence observed; this may be due to little training. However, in all other questions, users gave correct or approximate answers.
In questions regarding $T2$, users could identify and interpret the intersections correctly, as well as identify and analyze spurious intersections. The summary of the answers to the quantitative questions is available in Fig.~\ref{fig:user-study-results}:
In the following, we comment on some important aspects:

\myparagraph{Space representation (Q1, Q2, Q4):}
Most users found it easy to understand the representation of space, including the area of objects; those who found it more difficult agreed that it could be solved with proper training. One participant commented, \emph{``It is possible to easily visualize distances across the curve path and how objects have changed over time and visualize their region in space and intersection with other objects"}. Another comment: \emph{``The most complex thing for me is understanding the transformation from 2D into 1D. With this knowledge, exploration becomes simple."}

\myparagraph{Representation of intersections (Q5, Q6, Q7, Q9)}:
Seven users agreed that it was easy to identify intersections; one commented, \emph{``Intersections are clear in the visualization"}. Six of the users found it easy to identify spurious intersections ($Q6$). The participant who strongly disagreed with $Q6$ commented, \emph{``If you do not use the selection and the Intersection view, it is hardly possible to differentiate between spurious and non-spurious intersections"}. We agree that the MoReVis view does not have the necessary support to differentiate spurious intersections; however, we offer a rich set of interactive features to address this limitation. Five participants strongly agree that the representation of intersections makes it easier to interpret the data ($Q9$), with two agreeing. One commented, \emph{``It helps to visualize information that would be very chaotic with the data view alone"}.

\figUserStudyResults

\myparagraph{Auxiliary Views (Q2, Q3, Q7, Q8):}
Beyond the MoReVis technique, users also considered the other interface tools necessary. Despite the simple blending applied in the color bar, the response ($Q3$) was positive. In the comments on question $Q2$, users mentioned this graphic as helpful for representing space. One of the comments was, \emph{``By looking at the color bar and examining how the curves in the MoReVis view translate to the original data view, one can understand where each original region was mapped"}. Similarly, the intersection view was rated as necessary as well. Most participants found the selection (to highlight the Intersection View) very useful; one of the comments in $Q7$ was, \emph{``The intersection view helps"}. When asked about the most essential linked views, all users responded that the Intersection View and the Data View were necessary, while only two found the parallel coordinates important. This could have occurred due to the small size of the datasets.

\myparagraph{Scope for improvements:}
Although MoReVis showed promising results in representing spatiotemporal features of objects, there are areas where the technique can be improved. 
Three of the eight participants claimed intermediate experience with visualization, and one claimed advanced knowledge. Because of this, it was suggested improvements to our proposal, particularly regarding the visualization interface. Some users suggested adding more interactions with the Data View, a play button to show the data as an animation, and a link to the Intersection View, showing the respective intersection and data while the animation occurs.
A suggestion was also to permit queries to compare pairs of trajectories, comparing the distance and trends between two highlighted objects.
One user also suggested higher-level functionality incorporating anomaly detection techniques to indicate significant aspects of the data to the user. 
These functionalities could be added to MoReVis with little effort, and we consider them possible future work to be pursued in this line of research.

\section{Discussion}

As shown in Sec.~\ref{sec:uses}, MoReVis can provide an effective spatiotemporal visual summary of moving regions datasets through its representation of space and the areas and intersections.
This section discusses relevant points concerning the MoReVis technique, its limitations, and opportunities for future improvements.

\looseness=-1
\myparagraph{Projection of centroids:}
In our current implementation, we assume that the region covered by an object at each timestep is convex.
When they are not convex, the centroid could fall outside the region.
Therefore, the spatial representation might not be accurate for some projection methods (such as PCA).
We highlight that this is not a strong constraint since, in this case, one can use the convex hull of each region or
use distances between the areas instead of the distance between centroids.
All the current experiments and use cases presented in this paper were done using convex moving regions. 
We plan to investigate how our algorithm and visualization would behave for non-convex moving regions in the future.

\myparagraph{Intersection representation:} 
To develop our optimization model for intersection representation, we chose a set of goals, as it is not always possible to obtain a perfect solution.
We opt to represent every intersection of the original space with the cost of over-representing intersections areas and spurious intersections.
Sec.~\ref{sec:evaluation} verified that these error values were small in the two example datasets.
Also, to deal with this error, our visualization interface added visual hints that indicate the timesteps and regions with the presence of errors.
Furthermore, the Intersection View permits a detailed verification of intersections presented by our approach.

\myparagraph{Scalability:}
Our solution solves a mixed-integer program for each time step, which is computationally expensive. 
For example, if we have $n$ objects, the optimization problem can have up to $n^2-n$ integer variables. 
In the case of the HURDAT dataset, we used only 70 hurricanes, and it took 5 seconds, but if we use all 298 objects, it will take a few minutes. 
Similarly, in the WILDTRACK dataset, we filtered 14 objects\,---\,people who had more permanence in the video scene\,---\,and our solution took 5 seconds; however, if we process the entire dataset (282 objects), it takes 5 minutes. 
We are interested in improving this process stage, which is an immediate future work. 
Nevertheless, this task can be pre-processed since it is only performed once. We used this strategy in the user study to have an interactive experience. 
We can also see scalability in the visual representation; MoReVis might be cluttered for large datasets. In this case, interactivity is essential to reduce the number of moving regions shown. We also plan to investigate the use of alternative representations (such as density-based approaches) to overcome this limitation.

\myparagraph{Future applications:} 
We plan to apply our method to analyze other types of datasets in the future. One possible application is to analyze clusters of trajectory data. It is easy to see that a group of trajectories can be seen as a moving region. Another application is the visualization of trajectories with uncertainty. These trajectories are often used to model the spatial uncertainty due, for example, to errors in sensor measurements or variations in prediction models. We believe that MoReVis can be effective in such applications.
One useful and simple adaptation would be to consider datasets where objects can present spatial splits and merges as the storm cells analyzed in Hornero~\cite{diehl2021hornero}.

\section{Conclusion}
\looseness=-1
We presented MoReVis, a visual summary that provides an overview of moving regions datasets. 
This technique is based on a carefully designed optimization problem to build the visualization layout.
MoReVis is applicable in several domains and analysis situations, as shown in our use cases and discussion. 
Our main directions for future work are to reduce the computation cost of our algorithm and apply our technique to summarize clustering results and trajectories that model the spatial uncertainty of moving objects.


 \acknowledgments{
This work was supported by CNPq-Brazil (grant \#311144/2022-5), Carlos Chagas Filho Foundation for Research Support of Rio de Janeiro State (FAPERJ)-Brazil (grant \#E-26/201.424/2021), S\~ao Paulo Research Foundation (FAPESP)-Brazil (grant \#2021/07012-0), and the School of Applied Mathematics at Fundação Getulio Vargas (FGV/EMAp). 
Any opinions, findings, conclusions, or recommendations expressed in this material are those of the authors and do not necessarily reflect the views of the CNPq, FAPESP, FAPERJ, or FGV.
}

\bibliographystyle{abbrv-doi}

\bibliography{paper}

\begin{thebibliography}{10}

\bibitem{AIGNER:2011:VISUALIZATION}
W.~Aigner, S.~Miksch, H.~Schumann, and C.~Tominski.
\newblock {\em Visualization of Time-Oriented Data}.
\newblock Springer Science \& Business Media, 2011.

\bibitem{Andrienko2013Visual}
N.~Andrienko and G.~Andrienko.
\newblock Visual analytics of movement: An overview of methods, tools and
  procedures.
\newblock {\em Information Visualization}, 12(1):3--24, 2013.

\bibitem{andrienko2013space}
N.~Andrienko, G.~Andrienko, L.~Barrett, M.~Dostie, and P.~Henzi.
\newblock Space {T}ransformation for {U}nderstanding {G}roup {M}ovement.
\newblock {\em IEEE TVCG}, 19(12):2169--2178, 2013.

\bibitem{Arendt2017They}
D.~Arendt and M.~Pirrung.
\newblock The ``y'' of it {M}atters, {E}ven for {S}toryline {V}isualization.
\newblock In {\em 2017 IEEE Conference on Visual Analytics Science and
  Technology (VAST)}, pp. 81--91, 2017.

\bibitem{Ayesha2020}
S.~Ayesha, M.~K. Hanif, and R.~Talib.
\newblock Overview and comparative study of dimensionality reduction techniques
  for high dimensional data.
\newblock {\em Information Fusion}, 59:44--58, 2020.

\bibitem{bach2017descriptive}
B.~Bach, P.~Dragicevic, D.~Archambault, C.~Hurter, and S.~Carpendale.
\newblock A {D}escriptive {F}ramework for {T}emporal {D}ata {V}isualizations
  {B}ased on {G}eneralized {S}pace-{T}ime {C}ubes.
\newblock {\em Comput. Graph. Forum}, 36(6):36--61, sep 2017.

\bibitem{bach2014visualizing}
B.~Bach, E.~Pietriga, and J.-D. Fekete.
\newblock Visualizing dynamic networks with matrix cubes.
\newblock In {\em Proceedings of the SIGCHI Conference on Human Factors in
  Computing Systems}, CHI '14, pp. 877--886. Association for Computing
  Machinery, New York, NY, USA, 2014.

\bibitem{bauerle2022where}
A.~B{\"a}uerle, C.~van Onzenoodt, S.~der Kinderen, J.~J. Westberg,
  D.~J{\"o}nsson, and T.~Ropinski.
\newblock Where did my lines go? visualizing missing data in parallel
  coordinates.
\newblock {\em Computer Graphics Forum}, 41(3):235--246, 2022.

\bibitem{bonchi2009privacy}
F.~Bonchi.
\newblock Privacy preserving publication of moving object data.
\newblock In {\em Privacy in Location-Based Applications}, pp. 190--215.
  Springer, 2009.

\bibitem{Buchmuller:2018:MVCTST}
J.~Buchm{\"u}ller, D.~J{\"a}ckle, E.~Cakmak, U.~Brandes, and D.~A. Keim.
\newblock Motionrugs: {V}isualizing {C}ollective {T}rends in {S}pace and
  {T}ime.
\newblock {\em IEEE TVCG}, 25(1):76--86, 2019.

\bibitem{spatialrugs}
J.~F. Buchm{\"u}ller, U.~Schlegel, E.~Cakmak, D.~A. Keim, and E.~Dimara.
\newblock Spatialrugs: A compact visualization of space and time for analyzing
  collective movement data.
\newblock {\em Computers \& Graphics}, 101:23--34, 2021.

\bibitem{camargo2007cluster}
S.~J. Camargo, A.~W. Robertson, S.~J. Gaffney, P.~Smyth, and M.~Ghil.
\newblock Cluster {A}nalysis of {T}yphoon {T}racks. {P}art {I}: {G}eneral
  {P}roperties.
\newblock {\em Journal of Climate}, 20(14):3635 -- 3653, 2007.

\bibitem{Chavdarova2018Wildtrack}
T.~Chavdarova, P.~Baqu{\'e}, S.~Bouquet, A.~Maksai, C.~Jose, T.~Bagautdinov,
  L.~Lettry, P.~Fua, L.~Van~Gool, and F.~Fleuret.
\newblock {WILDTRACK: A Multi-Camera HD Dataset for Dense Unscripted Pedestrian
  Detection}.
\newblock In {\em Proceedings of the IEEE Conference on Computer Vision and
  Pattern Recognition (CVPR)}, June 2018.

\bibitem{chen2015survey}
W.~Chen, F.~Guo, and F.-Y. Wang.
\newblock {A Survey of Traffic Data Visualization}.
\newblock {\em IEEE Transactions on Intelligent Transportation Systems},
  16(6):2970--2984, 2015.

\bibitem{Cox2013Visualizing}
J.~Cox, D.~House, and M.~Lindell.
\newblock Visualizing uncertainty in predicted hurricane tracks.
\newblock {\em International Journal for Uncertainty Quantification},
  3:143--156, 01 2013.

\bibitem{van2008visualizing}
L.~der Maaten and G.~Hinton.
\newblock {Visualizing data using t-SNE.}
\newblock {\em Journal of machine learning research}, 9(11), 2008.

\bibitem{diehl2021hornero}
A.~Diehl, R.~Pelorosso, J.~Ruiz, R.~Pajarola, M.~E. Gr{\"o}ller, and
  S.~Bruckner.
\newblock Hornero: Thunderstorms characterization using visual analytics.
\newblock {\em Computer Graphics Forum}, 40(3):299--310, 2021.

\bibitem{dixon1993titan}
M.~Dixon and G.~Wiener.
\newblock Titan: Thunderstorm identification, tracking, analysis, and
  nowcasting---a radar-based methodology.
\newblock {\em Journal of Atmospheric and Oceanic Technology}, 10(6):785 --
  797, 1993.

\bibitem{domingo2012microaggregation}
J.~Domingo-Ferrer and R.~Trujillo-Rasua.
\newblock Microaggregation-and permutation-based anonymization of movement
  data.
\newblock {\em Information Sciences}, 208:55--80, 2012.

\bibitem{Elzen2014Dynamic}
S.~v.~d. Elzen, D.~Holten, J.~Blaas, and J.~J. van Wijk.
\newblock {Dynamic Network Visualization withExtended Massive Sequence Views}.
\newblock {\em IEEE TVCG}, 20(8):1087--1099, 2014.

\bibitem{ferreira2013vector}
N.~Ferreira, J.~T. Klosowski, C.~E. Scheidegger, and C.~T. Silva.
\newblock Vector field k-means: Clustering trajectories by fitting multiple
  vector fields.
\newblock {\em Computer Graphics Forum}, 32(3pt2):201--210, 2013.

\bibitem{evaluating2020filho}
J.~Filho, W.~Stuerzlinger, and L.~Nedel.
\newblock Evaluating an immersive space-time cube geovisualization for
  intuitive trajectory data exploration.
\newblock {\em IEEE TVCG}, 26(01):514--524, jan 2020.

\bibitem{fonseca2021system}
C.~M. Fonseca and J.~G.~S. Paiva.
\newblock A system for visual analysis of objects behavior in surveillance
  videos.
\newblock In {\em 2021 34th SIBGRAPI Conference on Graphics, Patterns and
  Images (SIBGRAPI)}, pp. 176--183, 2021.

\bibitem{1dordering}
M.~Franke, H.~Martin, S.~Koch, and K.~Kurzhals.
\newblock {Visual Analysis of Spatio-temporal Phenomena with 1D Projections}.
\newblock {\em Computer Graphics Forum}, 40(3):335--347, 2021.

\bibitem{GUO:2006:SPATIAL}
D.~Guo and M.~Gahegan.
\newblock Spatial ordering and encoding for geographic data mining and
  visualization.
\newblock {\em Journal of Intelligent Information Systems}, 27:243--266, 11
  2006.

\bibitem{Harrower:2007:Cognitive}
M.~Harrower.
\newblock {The Cognitive Limits of Animated Maps}.
\newblock {\em Cartographica: The International Journal for Geographic
  Information and Geovisualization}, 42(4):349--357, 2007.

\bibitem{hoeferlin2013interactive}
M.~H{\"o}ferlin, B.~H{\"o}ferlin, G.~Heidemann, and D.~Weiskopf.
\newblock {Interactive Schematic Summaries for Faceted Exploration of
  Surveillance Video}.
\newblock {\em IEEE Transactions on Multimedia}, 15(4):908--920, 2013.

\bibitem{inselberg1990parallel}
A.~Inselberg and B.~Dimsdale.
\newblock {Parallel Coordinates: A Tool For Visualizing Multi-Dimensional
  Geometry}.
\newblock In {\em Proceedings of the First IEEE Conference on Visualization:
  Visualization `90}, pp. 361--378, 1990.

\bibitem{joshi2012survey}
K.~A. Joshi and D.~G. Thakore.
\newblock {A Survey on Moving Object Detection and Tracking in Video
  Surveillance System}.
\newblock {\em International Journal of Soft Computing and Engineering},
  2(3):44--48, 2012.

\bibitem{spacetimecube}
M.~Kraak.
\newblock The space-time cube revisited from a geovisualization perspective.
\newblock In {\em ICC 2003 : Proceedings of the 21st International Cartographic
  Conference}, pp. 1988--1996. International Cartographic Association, New
  Zealand, 2003.
\newblock 21st international Cartographic Conference, ICC 2003 : Cartographic
  renaissance, ICC ; Conference date: 10-08-2003 Through 16-08-2003.

\bibitem{kruskal1964multidimensional}
J.~B. Kruskal.
\newblock {Multidimensional Scaling by Optimizng Goodness of a Fit to Nonmetric
  Hypothesis}.
\newblock {\em Psychometrika}, 29(1):1--27, 1964.

\bibitem{hurdat}
C.~Landsea and J.~Beven.
\newblock { The revised Atlantic hurricane database (HURDAT2) }, howpublished =
  {\url{https://www.aoml.noaa.gov/hrd/hurdat}}, note = {Accessed: 2021-11-17}.

\bibitem{lee2011mixed}
J.~Lee and S.~Leyffer.
\newblock {\em Mixed integer nonlinear programming}, vol. 154.
\newblock Springer Science \& Business Media, 2011.

\bibitem{lee2019space}
T.-Y. Lee and K.~Wittenburg.
\newblock {Space-Time Slicing: Visualizing Object Detector Performance in
  Driving Video Sequences}.
\newblock In {\em 2019 IEEE Pacific Visualization Symposium (PacificVis)}, pp.
  318--322, 2019.

\bibitem{lisecure2021}
H.~Li, T.~Xiezhang, C.~Yang, L.~Deng, and P.~Yi.
\newblock {Secure Video Surveillance Framework in Smart City}.
\newblock {\em Sensors}, 21(13), 2021.

\bibitem{Li2011MoveMine}
Z.~Li, J.~Han, M.~Ji, L.-A. Tang, Y.~Yu, B.~Ding, J.-G. Lee, and R.~Kays.
\newblock {MoveMine: Mining Moving Object Data for Discovery of Animal Movement
  Patterns}.
\newblock {\em ACM Trans. Intell. Syst. Technol.}, 2(4), jul 2011.

\bibitem{Liu2013Storyflow}
S.~Liu, Y.~Wu, E.~Wei, M.~Liu, and Y.~Liu.
\newblock {StoryFlow: Tracking the Evolution of Stories}.
\newblock {\em IEEE TVCG}, 19(12):2436--2445, 2013.

\bibitem{lu1993spatial}
H.~Lu and B.~C. Ooi.
\newblock {Spatial Indexing: Past and Future}.
\newblock {\em {IEEE} Data Eng. Bull.}, 16(3):16--21, 1993.

\bibitem{mcinnes2020umap}
L.~McInnes, J.~Healy, N.~Saul, and L.~Gro{\ss}berger.
\newblock {UMAP: Uniform Manifold Approximation and Projection}.
\newblock {\em Journal of Open Source Software}, 3(29):861, 2018.

\bibitem{meghdadi2013interactive}
A.~H. Meghdadi and P.~Irani.
\newblock {Interactive Exploration of Surveillance Video through Action Shot
  Summarization and Trajectory Visualization}.
\newblock {\em IEEE TVCG}, 19(12):2119--2128, 2013.

\bibitem{mirzargar2014curve}
M.~Mirzargar, R.~T. Whitaker, and R.~M. Kirby.
\newblock {Curve Boxplot: Generalization of Boxplot for Ensembles of Curves}.
\newblock {\em IEEE TVCG}, 20(12):2654--2663, 2014.

\bibitem{pearson1901liii}
K.~Pearson.
\newblock Liii. on lines and planes of closest fit to systems of points in
  space.
\newblock {\em The London, Edinburgh, and Dublin philosophical magazine and
  journal of science}, 2(11):559--572, 1901.

\bibitem{pena2021hyperstorylines}
V.~Pe{\~n}a-Araya, T.~Xue, E.~Pietriga, L.~Amsaleg, and A.~Bezerianos.
\newblock {HyperStorylines: Interactively Untangling Dynamic Hypergraphs}.
\newblock {\em Information Visualization}, 21(1):38--62, 2022.

\bibitem{raty2010survey}
T.~D. R{\"a}ty.
\newblock {Survey on Contemporary Remote Surveillance Systems for Public
  Safety}.
\newblock {\em IEEE Transactions on Systems, Man, and Cybernetics, Part C
  (Applications and Reviews)}, 40(5):493--515, 2010.

\bibitem{2022-JamVis}
E.~Rodriguez, N.~Ferreira, and J.~Poco.
\newblock Jamvis: Exploration and visualization of traffic jams.
\newblock {\em The European Physical Journal Special Topics (EPJ ST)}, 2022.

\bibitem{sss-gi2001}
G.~Smith, W.~Stuerzlinger, and T.~Salzman.
\newblock 3d scene manipulation with 2d devices and constraints.
\newblock In {\em Proceedings of the Graphics Interface 2001 Conference, June
  7-9 2001, Ottawa, Ontario, Canada}, pp. 135--142, June 2001.

\bibitem{Tanahashi2012Design}
Y.~Tanahashi and K.-L. Ma.
\newblock {Design Considerations for Optimizing Storyline Visualizations}.
\newblock {\em IEEE TVCG}, 18(12):2679--2688, 2012.

\bibitem{Tang2019istoryline}
T.~Tang, S.~Rubab, J.~Lai, W.~Cui, L.~Yu, and Y.~Wu.
\newblock {iStoryline: Effective Convergence to Hand-drawn Storylines}.
\newblock {\em IEEE TVCG}, 25:769--778, 2019.

\bibitem{improved:2003:tejada}
E.~Tejada, R.~Minghim, and L.~G. Nonato.
\newblock On improved projection techniques to support visual exploration of
  multi-dimensional data sets.
\newblock {\em Information Visualization}, 2(4):218--231, 2003.

\bibitem{thudt2013visits}
A.~Thudt, D.~Baur, and S.~Carpendale.
\newblock {Visits: A Spatiotemporal Visualization of Location Histories}.
\newblock In M.~Hlawitschka and T.~Weinkauf, eds., {\em EuroVis - Short
  Papers}. The Eurographics Association, 2013.

\bibitem{tripathi2016direction}
P.~K. Tripathi, M.~Debnath, and R.~Elmasri.
\newblock {A Direction Based Framework for Trajectory Data Analysis}.
\newblock In {\em Proceedings of the 9th ACM International Conference on
  Pervasive Technologies Related to Assistive Environments}, PETRA '16.
  Association for Computing Machinery, New York, NY, USA, 2016.

\bibitem{Valdivia2021Analyzing}
P.~Valdivia, P.~Buono, C.~Plaisant, N.~Dufournaud, and J.-D. Fekete.
\newblock {Analyzing Dynamic Hypergraphs with Parallel Aggregated Ordered
  Hypergraph Visualization}.
\newblock {\em IEEE TVCG}, 27(1):1--13, 2021.

\bibitem{van2016block}
T.~C. van Dijk, M.~Fink, N.~Fischer, F.~Lipp, P.~Markfelder, A.~Ravsky,
  S.~Suri, and A.~Wolff.
\newblock {Block Crossings in Storyline Visualizations}.
\newblock In Y.~Hu and M.~N{\"o}llenburg, eds., {\em Graph Drawing and Network
  Visualization}, pp. 382--398. Springer International Publishing, Cham, 2016.

\bibitem{wang2011interactive}
Z.~Wang, H.~Guo, B.~Yu, and X.~Yuan.
\newblock Interactive visualization of 160 years' global hurricane trajectory
  data.
\newblock In {\em Proceedings of the IEEE Pacific Visualization Symposium
  (Poster), Hong Kong}, pp. 37--38, 2011.

\bibitem{Wang2014Urban}
Z.~Wang and X.~Yuan.
\newblock {Urban Trajectory Timeline Visualization}.
\newblock In {\em 2014 International Conference on Big Data and Smart Computing
  (BIGCOMP)}, pp. 13--18, 2014.

\bibitem{ware2013information}
C.~Ware.
\newblock {\em Information Visualization: Perception for Design}.
\newblock Information Visualization: Perception for Design. Elsevier Science,
  2013.

\bibitem{stablevisualsummaries}
J.~Wulms, J.~Buchmuller, W.~Meulemans, K.~Verbeek, and B.~Speckmann.
\newblock {Stable Visual Summaries for Trajectory Collections}.
\newblock In {\em 2021 IEEE 14th Pacific Visualization Symposium (PacificVis)},
  pp. 61--70. IEEE Computer Society, Los Alamitos, CA, USA, apr 2021.

\end{thebibliography}
 

\vbox{%
\begin{wrapfigure}{l}{70pt}
{
	\includegraphics[width=1in,height=1.25in,clip,keepaspectratio]{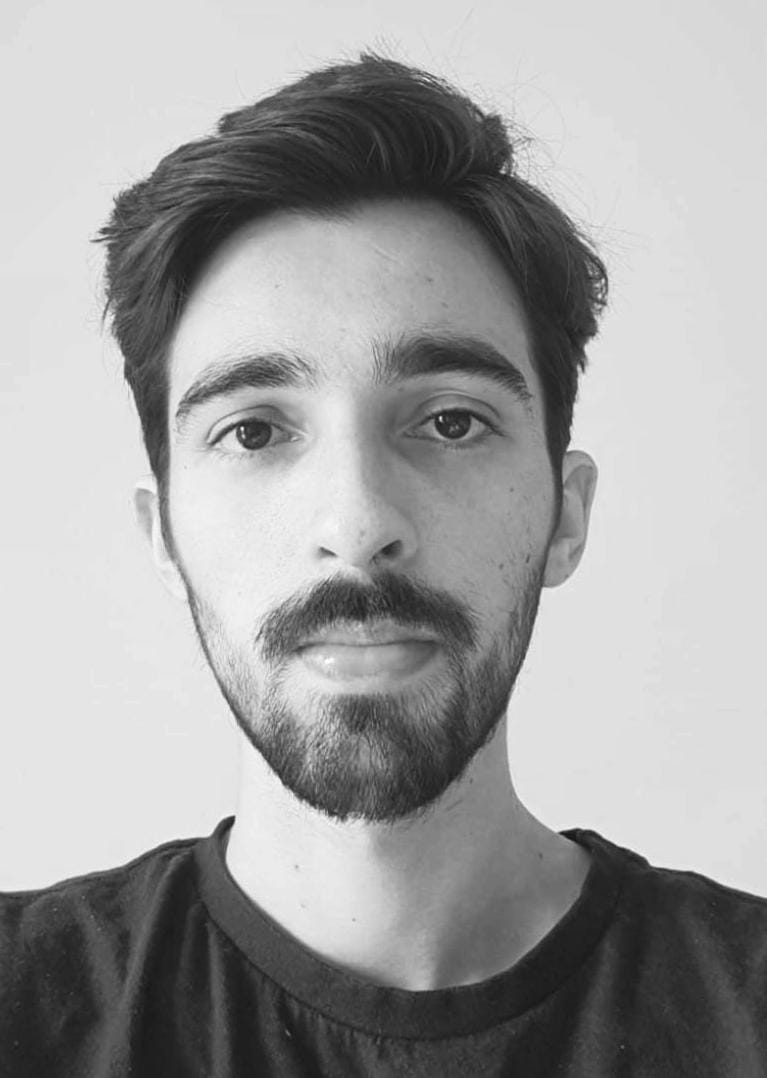}
}%
\end{wrapfigure}
\noindent\small 
\\
\\
\textbf{Giovani~Valdrighi}
  received his B.Sc. (2021) in Applied Mathematics from the School of Applied Mathematics of Fundação Getulio Vargas (FGV-EMAP), Rio de Janeiro, Brazil.
  His research interests include visualization,  machine learning, and data science.
  He is currently an M.Sc. candidate in Mathematical Modeling at FGV-EMAP.}
\vspace{0.5cm}

\vbox{%
\begin{wrapfigure}{l}{70pt}
{
\vspace*{20pt}
    \includegraphics[width=1in,height=1.25in,clip,keepaspectratio]{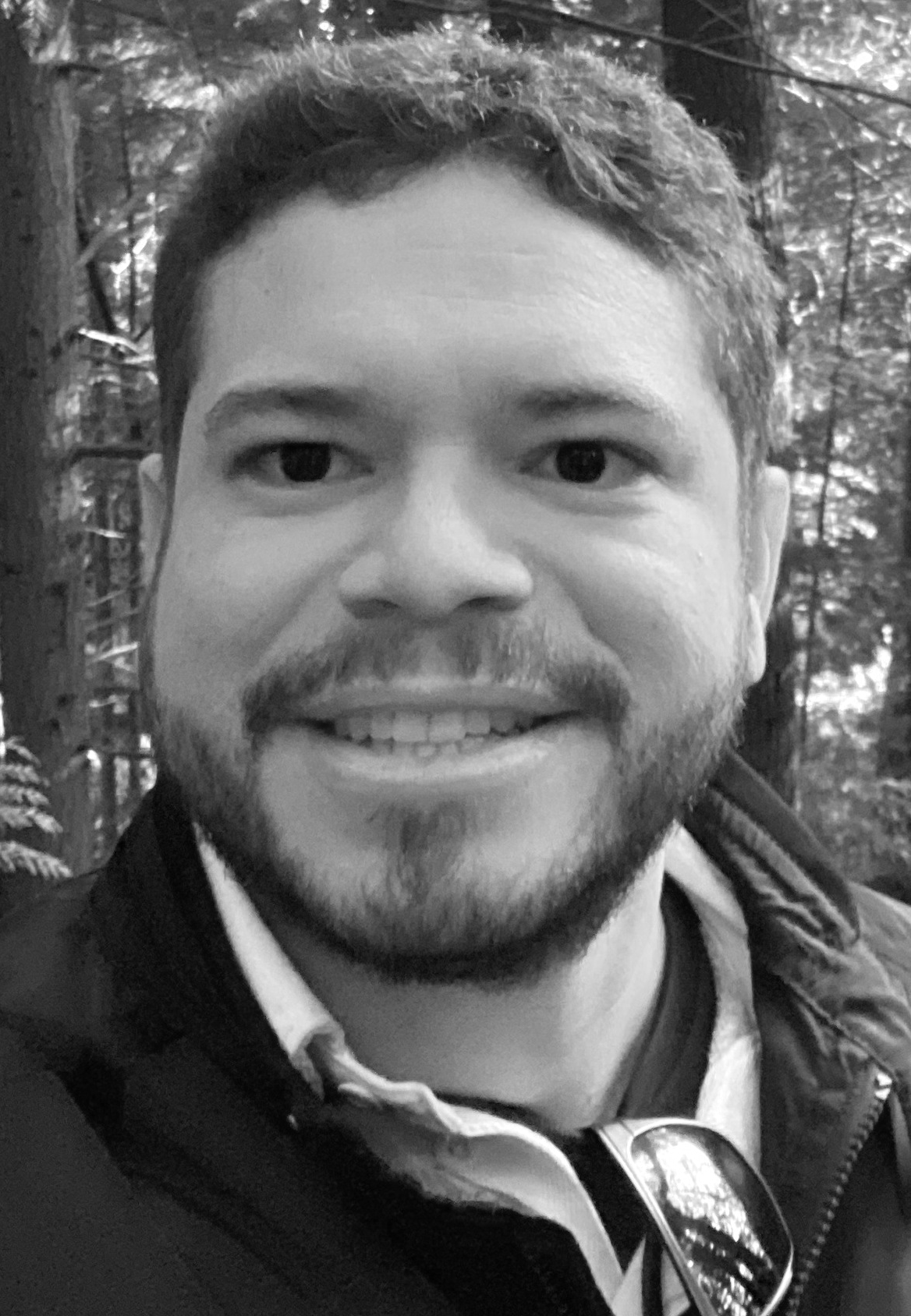}
	\vspace*{15pt}
}%
\end{wrapfigure}
\noindent\small 
\\
\\
\\
\textbf{Nivan~Fereira}
  is an Assistant Professor at Centro de Informática at Universidade Federal de Pernambuco in Brazil. 
  He received his Ph.D. in Computer Science in 2015 from New York University, his M.Sc. in Mathematics in 2010, and B.Sc. degree in Computer Science both from Universidade Federal de Pernambuco.
  His research focuses on many aspects of interactive data visualization, in particular systems and techniques for analysis spatiotemporal datasets.
}

\vspace{0.5cm}

\vbox{%
\begin{wrapfigure}{l}{70pt}
{
\vspace*{13pt}
	\includegraphics[width=1in,height=1.25in,clip,keepaspectratio]{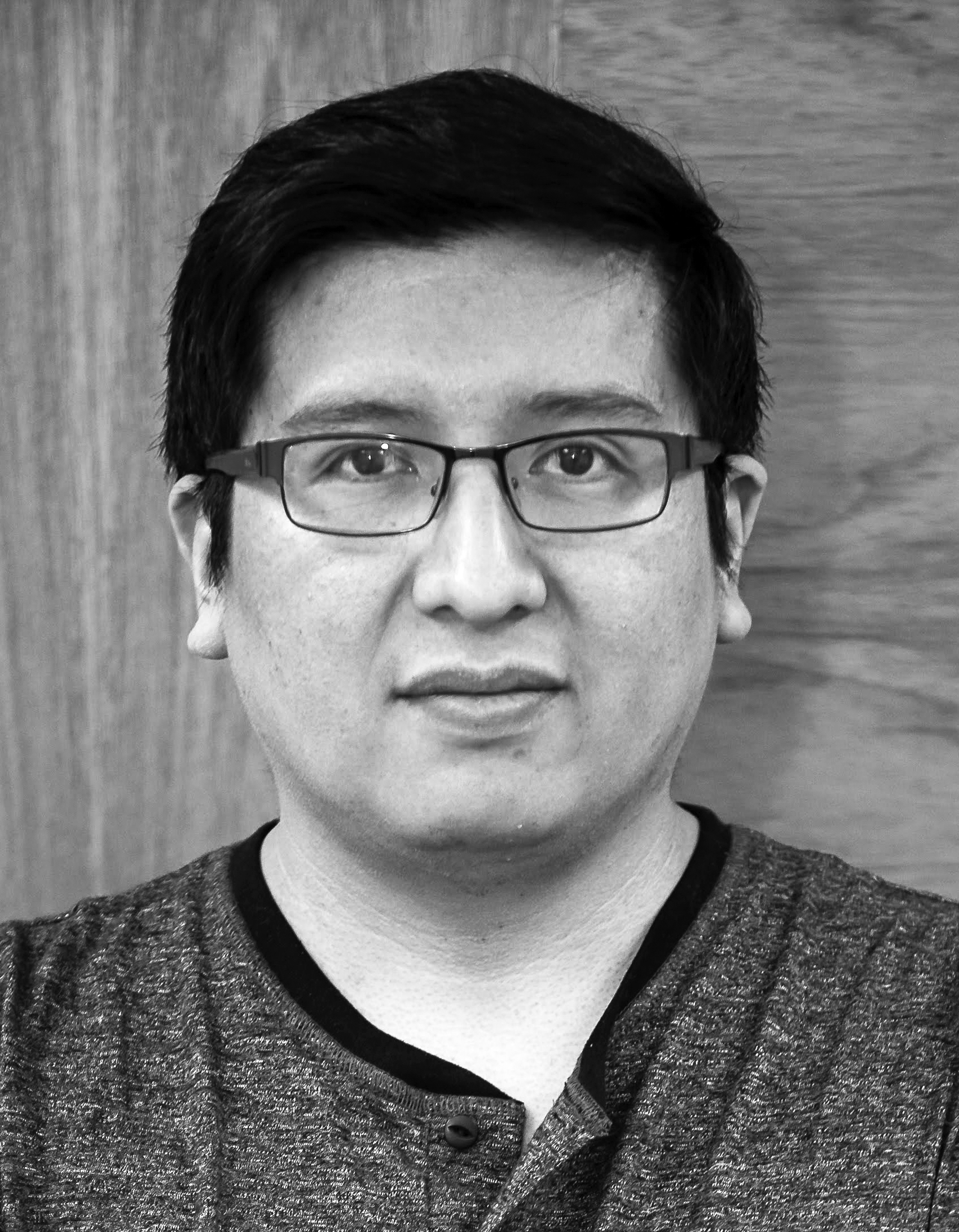}
}%
\end{wrapfigure}
\noindent\small 
\\
\\
\\
\textbf{Jorge~Poco}
is an Associate Professor at the School of Applied Mathematics at Fundação Getulio Vargas Rio de Janeiro-Brazil. He received his Ph.D. in Computer Science in 2015 from New York University, his M.Sc. in Computer Science in 2010 from the University of São Paulo (Brazil), and his B.E. in System Engineering in 2008 from the National University of San Agustín (Peru).
His research interests are data visualization, visual analytics, machine learning, and data science. He has served on several program committees, including IEEE SciVis, IEEE InfoVis, VAST, and EuroVis.}

\end{document}